%% file: Yadav-etal_manuscript.tex
\documentclass[12pt]{article}
\RequirePackage[OT1]{fontenc}
\usepackage{amsthm,amsmath,amssymb,amsfonts,bm,enumerate,xcolor,graphicx,natbib,color,url}
\RequirePackage[colorlinks,citecolor=blue,urlcolor=blue]{hyperref}
\usepackage{hyperref}
\usepackage{subcaption} 
\usepackage{textcmds}
\usepackage{booktabs}
\usepackage{lipsum}
\usepackage{cancel}
\newcommand{\bs}{\bm{s}}

\allowdisplaybreaks
\input{macros.tex}

\input{teachingmacros.tex}

\input{packages.tex}

\def\bs{{\boldsymbol{s}}}

\newcommand{\blind}{1}

\begin{document}
\thispagestyle{empty}
\baselineskip=28pt
\vskip 5mm
\begin{center} {\Large{\bf Spatial hierarchical modeling of threshold exceedances using rate mixtures}}
\end{center}

\baselineskip=12pt

\vskip 5mm

\if1\blind
{
\begin{center}
\large
Rishikesh Yadav$^1$, Rapha\"el Huser$^1$ and Thomas Opitz$^2$\\ 
\end{center}

\footnotetext[1]{
\baselineskip=10pt Computer, Electrical and Mathematical Sciences and Engineering (CEMSE) Division, King Abdullah University of Science and Technology (KAUST), Thuwal 23955-6900, Saudi Arabia. E-mails: rishikesh.yadav@kaust.edu.sa; raphael.huser@kaust.edu.sa}
\footnotetext[2]{
\baselineskip=10pt  INRAE, UR546 Biostatistics and Spatial Processes, 228, Route de l'A\'erodrome, CS 40509, 84914 Avignon, France. E-mail: thomas.opitz@inra.fr}
} \fi

\baselineskip=20pt
\vskip 4mm
\centerline{\today}
\vskip 6mm

{\large{\bf Abstract}}
We develop new flexible univariate models for light-tailed and heavy-tailed data, which extend a hierarchical representation of the generalized Pareto (GP) limit for threshold exceedances. These models can accommodate departure from asymptotic threshold stability in finite samples while keeping the asymptotic GP distribution as a special (or boundary) case and can capture the tails and the bulk jointly without losing much flexibility. Spatial dependence is modeled through a latent process, while the data are assumed to be conditionally independent. Focusing on a gamma-gamma model construction, we design penalized complexity priors for crucial model parameters, shrinking our proposed spatial Bayesian hierarchical model toward a simpler reference whose marginal distributions are GP with moderately heavy tails. Our model can be fitted in fairly high dimensions using Markov chain Monte Carlo by exploiting the Metropolis-adjusted Langevin algorithm (MALA), which guarantees fast convergence of Markov chains with efficient block proposals for the latent variables. We also develop an adaptive scheme to calibrate the MALA tuning parameters. Moreover, our model avoids the expensive numerical evaluations of multifold integrals in censored likelihood expressions. We demonstrate our new methodology by simulation and application to a dataset of extreme rainfall events that occurred in Germany. Our fitted gamma-gamma model provides a satisfactory performance and can be successfully used to predict rainfall extremes at unobserved locations.


\baselineskip=20pt

\par\vfill\noindent



\newpage
\baselineskip=26pt
\section{Introduction}
Atmospheric, meteorologic, hydrologic, and land surface processes among others are prone to extreme episodes, which are believed to increase in frequency and severity in the context of a changing climate and global warming \citep[see, e.g.,][]{Witze:2018,Power.Delage:2019,Arneth.etal:2019}.
	To quantify environmental risk, e.g., associated with weather variables \citep{Davison.Gholamrezaee:2012,Huser.Davison:2014a,deFondeville.Davison:2018} or air pollution \citep{Eastoe.Tawn:2009,Vettori.etal:2019,Vettori.etal:2020}, or to attribute certain extreme events to human influence \citep{Risser.Wehner:2017}, it is crucial to model and predict the univariate behavior of extreme events, while accounting for spatial dependence. Extreme-Value Theory (EVT) provides a natural methodological framework to tackle this problem; for more details, see, e.g., the review papers \citet{Davison.etal:2012}, \citet{Davison.Huser:2015} and \citet{Davison.etal:2019}. 
	
In this paper, we first study a hierarchical construction that leads to new univariate tail models, which extend the classical EVT approach by gaining flexibility at finite levels. We then exploit this hierarchical construction in a Bayesian framework for the modeling of spatial extremes by embedding a latent process with spatial dependence, while keeping the desired unconditional marginal distributions.

Under mild conditions, classical univariate EVT suggests using the generalized Pareto (GP) distribution for modeling extreme events defined as high threshold exceedances \citep{Davison.Smith:1990}. More precisely, let $Y$ be a random variable following a distribution $F$  with 
finite or infinite upper endpoint $y_F=\sup\{y\in\Real: F(y)<1\}$. Then, for a wide class of distributions $F$, high 
threshold exceedances $(Y-u)\mid Y>u$ may be \emph{asymptotically} approximated as
\begin{equation}
\label{eq:GP}
\pr(Y-u\leq y\mid Y>u)={F(u+y)-F(u)\over 1-F(u)}\approx H_{\tau,\xi}(y)=1-(1+\xi y/\tau)^{-1/\xi},
\end{equation}
as the threshold $u$ converges to $y_F$, where $H_{\tau,\xi}(y)$ denotes the GP distribution function with scale parameter $\tau>0$ and shape parameter $\xi\in\Real$ (also called tail index), defined over $\{y>0:1+\xi y/\tau>0\}$. In other words, $1-F(y)\approx 
\zeta_u\{1-H_{\tau,\xi}(y-u)\}$, for $y>u$ large, where $\zeta_u=\text{Pr}(Y>u)$. When $\xi=0$, the distribution \eqref{eq:GP} is interpreted as the limit $\xi\to0$, and we obtain the exponential distribution function $H_{\tau,0}(y)=1-\exp(-y/\tau)$, $y>0$. When $\xi<0$, the support is bounded, whereas when $\xi\geq0$, it is unbounded. Short, light, and heavy tails correspond to $\xi<0$, $\xi=0$ and $\xi>0$, respectively, with $\xi$ controlling the tail weight.

In practice, the choice of a good threshold $u$ should reflect the transition around which the asymptotic regime takes place for the tail approximation \eqref{eq:GP} to be valid. This implies a bias-variance trade-off, as a high threshold $u$ leads to a good approximation (low bias) but yields a small number of exceedances (high variance), and vice versa for a low threshold. Experience shows that automatic threshold selection procedures are not always reliable. It is often difficult to find a good, natural and interpretable threshold, and parameter estimates are often sensitive to this choice \citep{Scarrott.MacDonald:2012}. This has motivated the development of \emph{sub-asymptotic} models for extremes, which are more flexible than the asymptotic GP distribution at finite levels, while keeping a GP-like behavior in the tail; see, e.g., \citet{Frigessi.etal:2003}, \citet{Carreau.Bengio:2009}, and \citet{Papastathopoulos.Tawn:2013}, among others; see \citet{Scarrott.MacDonald:2012} for a review of models describing jointly the bulk and the tail of the distribution. With sub-asymptotic tail models, parameter estimates are usually less sensitive to the threshold, the choice of which then becomes less crucial for inference, and we can thus set lower thresholds. In some approaches \citep[see, e.g.,][]{Naveau.etal:2016,Stein:2020a,Stein:2020b}, the threshold choice is even bypassed by adding parameters that provide separate control over bulk properties of the distribution, such that models are expected to provide a satisfactory fit of the tail, even if the whole sample is used for estimation. 


In this paper, we propose a novel Bayesian hierarchical modeling framework for sub-asymptotic threshold exceedances. It has an intuitive interpretation, permits fully Bayesian inference, and can naturally incorporate covariate information and be extended to the spatial setting. Several Bayesian hierarchical models were already proposed in the literature to model threshold exceedances; see, e.g., \citet{cooley2007bayesian}, \citet{opitz2018inla} and \citet{CastroCamilo.etal:2019}. However, unlike the latent Gaussian models proposed therein, our construction makes sure that the unconditional distribution (obtained after integrating out latent random effects) remains of the desired form, with the GP distribution as a particular case. 
Precisely, we extend the characterization of the GP distribution as an exponential mixture with rate parameter following a gamma distribution. Let ${\rm Exp}(\lambda)$ denote the exponential distribution with rate $\lambda>0$, ${\rm Gamma}(\alpha,\beta)$ denote the gamma distribution with rate $\alpha>0$ and shape $\beta>0$, i.e., with density $g(y)=\{\Gamma(\beta)\}^{-1}\alpha^{\beta} y^{\beta-1}\exp(-\alpha y)$, $y>0$, and ${\rm GP}(\tau,\xi)$ denote the GP distribution with scale $\tau>0$ and shape $\xi$ as defined in \eqref{eq:GP}. Then we have
\begin{equation}
\label{eq:GP.mixture}
\left.\begin{array}{rl}
Y\mid \Lambda &\sim {\rm Exp}(\Lambda)\\
\Lambda&\sim {\rm Gamma}(\alpha,\beta)
\end{array}\right\}\Rightarrow Y\sim {\rm GP}(\alpha/\beta,1/\beta);  
\end{equation}
see \citet{Reiss.Thomas:2007}, \citet{Bortot.Gaetan:2014}, \citet{Bortot.Gaetan:2016}, \citet{Bopp.Shaby:2017} and \cite{Bacro.etal:2020}. In other words, exponentially-decaying tails become heavier by making their rate parameter $\Lambda$ random. 
By integrating out the latent variable $\Lambda$ in the hierarchical construction in \eqref{eq:GP.mixture}, we obtain the GP distribution for the data $Y$. Our new tail models (detailed further in \S\ref{sec:models} below) are constructed as in \eqref{eq:GP.mixture}, but we modify the top and/or lower levels of the hierarchy in order to gain in flexibility, while keeping the GP distribution with $\xi \geq 0$ as a special or boundary case. Moreover, we penalize departure from the GP distribution in the Bayesian framework by specifying penalized complexity (PC) priors \citep{Simpson.etal:2017} designed to shrink complex models toward simpler counterparts, thus preventing overfitting. This avoids estimating unreasonable tail models, and guarantees that the fitted distribution will not be too far away from the GP distribution which is supported by asymptotic theory, unless the data provide strong evidence that a different sub-asymptotic behavior prevails. In other words, our proposed models are constrained to remain in the \qq{neighborhood} of moderately heavy-tailed GP distributions. Our modeling approach based on extensions of \eqref{eq:GP.mixture} is general and can potentially generate a wide variety of new models with light and heavy tails and various behaviors in the bulk. Below, we mainly focus on a parsimonious extension of \eqref{eq:GP.mixture}, which assumes a gamma distribution in both levels of the hierarchy, although we also discuss other possible models with interesting tail properties.

For spatial modeling, we incorporate spatial dependence at the latent level, while assuming that the data are conditionally independent given the latent process. Specifically, we assume that the observed spatial process $Y(\bs)$, $\bs\in\calS\subset\Real^2$, may be described analogously to \eqref{eq:GP.mixture} with a hierarchical representation in terms of a latent spatially structured process $\Lambda(\bs)$, such that the data $Y(\bs_1)$ and $Y(\bs_2)$ at any two distinct locations $\bs_1,\bs_2\in\calS$, $\bs_1\neq\bs_2$, are independent given $\Lambda(\bs_1)$ and $\Lambda(\bs_2)$. This conditional independence assumption is common in Bayesian hierarchical models \citep{Banerjee.etal:2014, cooley2007bayesian, opitz2018inla} and is, to some extent, akin to using a ``nugget effect'' in classical geostatistics \citep{Cressie:1993}, which captures measurement errors or unstructured local variations in the data. Here, we make this assumption mainly to keep the model simple and identifiable, and for computational convenience in order to efficiently handle the censoring of non-extreme values in our inference procedure. More precisely, to fit our models to threshold excesses $Y(\bs)>u(\bs)$ for some moderately high threshold $u(\bs)$, we design a generic Markov chain Monte Carlo (MCMC) sampler that efficiently exploits the hierarchical representation. Low values such that $Y(\bs)\leq u(\bs)$ are treated as censored and imputed by simulation in our MCMC algorithm. Thanks to the conditional independence assumption, multivariate censoring 
can be conveniently reduced to univariate site-by-site imputations. This computational benefit is significant, and it contrasts with the high computational burden due to multivariate censoring in peaks-over-threshold inference for most spatial extremes models \citep{Wadsworth.Tawn:2014,Huser.etal:2017,Huser.Wadsworth:2019,castro2019local}. This approach allows us to tackle higher spatial dimensions more easily. Furthermore, to efficiently sample from the posterior distribution and accelerate the mixing of MCMC chains, we update the large number of latent parameters jointly (in one block) based on the Metropolis-adjusted Langevin algorithm (MALA, see \citet{Roberts.Tweedie:1996} and \citet{roberts1998optimal}) and we 
adaptively calibrate the MALA tuning parameters to obtain suitable acceptance rates. Our fully Bayesian algorithm can easily handle missing values and be used for prediction at unobserved locations. Finally, in spite of the conditional independence assumption, non-trivial extremal dependence structures may be obtained by carefully specifying the dependence structure of the latent process $\Lambda(\bm s)$. This contrasts with classical latent Gaussian  models for extremes \citep[e.g.,][]{cooley2007bayesian,opitz2018inla}, which are limited for capturing strong extremal dependence.


The paper is organized as follows. In \S\ref{sec:models}, we define our hierarchical construction of univariate distributions and characterize their tail behavior. Spatial hierarchical modeling is developed in \S\ref{sec:spatial}, and we describe Bayesian inference and our MCMC implementation using latent variables in \S\ref{sec:bayesian}. An extensive simulation study is reported in \S\ref{sec:simulation} showing that our approach works well in various scenarios. We use our approach based on a gamma-gamma hierarchical model to study extreme events in daily precipitation measurements recorded at $150$ sites in Germany in \S\ref{sec:application}. Concluding remarks are given in \S\ref{sec:conclusion}.

\section{Hierarchical models for threshold exceedances}
\label{sec:models}

\subsection{Univariate tail properties in rate mixture constructions}
For flexible sub-asymptotic tail modeling, we  seek to replace the distributions in the hierarchical representation \eqref{eq:GP.mixture} of the GP distribution by more general parametric families that contain the exponential-gamma mixture considered by \citet{Bopp.Shaby:2017} as a special (or boundary) case.

Specifically, we construct new rate mixture models for the data $Y\sim F$ as follows. We consider a family of distributions $F_{Y\mid \Lambda=\lambda}(\cdot)$ with rate parameter $\lambda$ and supported in $[0,+\infty)$, where $\Lambda\geq 0$ is  a  latent random variable, such that $Y\mid \Lambda\sim F_{Y\mid \Lambda}(\cdot)$. Equivalently, we have the following ratio representation, which is useful for simulation and inference: 
\begin{equation}\label{eq:ratio}
Y\mid \Lambda \stackrel{D}{=} {\tilde{Y}\over\Lambda},  \quad\mbox{with}\quad \Lambda\geq 0 \indep \tilde{Y}\geq 0,\quad  \tilde{Y}\sim F_{Y\mid \Lambda=1}(\cdot),
\end{equation}
where $ \text{``}\stackrel{D}{=} \text{"}$ means equality in distribution and $\indep$ denotes the independence of random variables.
The (unconditional) upper tail behavior of $Y$ is determined by the interplay between the upper tail  of $\tilde{Y}$ and the lower tail of  $\Lambda$ (i.e., the upper tail of $1/\Lambda$). We now shortly discuss two general and particularly interesting scenarios.  Recall that a positive function $r(\cdot)$ is regularly varying at infinity with index $a\in\mathbb{R}$ if $r(tx)/r(t)\to x^{a}$ as  $t\to\infty$ and $x>0$; when $a=0,$ $r(\cdot)$ is slowly varying at infinity.

In the first scenario, we assume that $1/\Lambda$ in \eqref{eq:ratio} has power-law tail decay, i.e., its distribution is regularly varying with index $-a<0$, such that ${\rm Pr}(1/\Lambda>y)=r_0(y)y^{-a}$, $y>0$, and $r_0(\cdot)>0$ is a slowly varying function. If the distribution $F_{Y\mid\Lambda=1}(\cdot)$ in \eqref{eq:ratio} has a lighter upper tail than that of $1/\Lambda$,  such that $\mathbb E(\tilde{Y}^{a+\varepsilon})<\infty$ for some $\varepsilon>0$ with $\tilde{Y}\sim F_{Y\mid\Lambda=1}(\cdot)$, then Breiman's Lemma \citep{Breiman.1965} implies that 
\begin{equation}\label{eq:breiman}
1-F(y)=\Pr(Y> y) \sim \mathbb{E}(\tilde{Y}^{a})\, \Pr(1/\Lambda>y), \quad y\rightarrow\infty.
\end{equation}
Therefore, the heavier-tailed random factor $1/\Lambda$ in \eqref{eq:ratio} dominates the tail behavior of $Y$ in this case, while the lighter-tailed random factor $\tilde{Y}$ contributes to extreme survival probabilities only through a scaling factor. 

In the second scenario, we assume that both $\tilde{Y}$ and $1/\Lambda$ in \eqref{eq:ratio} have tails of Weibull type, which are lighter than power-law tails. Formally, we assume that there exist regularly varying functions $\tilde{r}$, $r_{\Lambda}$ (with any index of regular variation), rate parameters $\tilde{\alpha},\alpha_\Lambda>0$ and shape parameters (also referred to as Weibull indexes) $\tilde{\eta},\eta_\Lambda>0$ such that 
\begin{equation}\label{eq:weibull}
\Pr(\tilde{Y}>y)=\tilde{r}(y)\exp(-\tilde{\alpha}y^{\tilde{\eta}} ), \quad \Pr(1/\Lambda>y)=r_{\Lambda}(y)\exp(-\alpha_{\Lambda}y^{\eta_{\Lambda}} ). 
\end{equation}
Then, the variable $Y$ constructed as in \eqref{eq:ratio} also has a tail of Weibull type, with representation $\Pr(Y>y)=r_Y(y)\exp(-\alpha_Y y^{\eta_Y} )$ similarly to \eqref{eq:weibull}. Its Weibull index is $\eta_Y=(\tilde{\eta}\eta_\Lambda)/(\tilde{\eta}+\eta_\Lambda) <\min(\tilde{\eta},\eta_\Lambda)$, such that the tail of $Y$ always has a slower decay rate than that of each random factor $\tilde{Y}$ and $1/\Lambda$, while its rate parameter $\alpha_Y$ is given by
$$
\alpha_Y=\tilde{\alpha}^{1-b}\alpha_\Lambda^b\left\{\left({\alpha_\Lambda\over\tilde{\alpha}}\right)^b+\left({\tilde{\alpha}\over\alpha_\Lambda}\right)^{1-b}\right\},\quad b={\tilde{\eta}\over\tilde{\eta}+\eta_\Lambda};
$$
see \citet{Arendarczyk.Debicki.2011}. In the following sections, \S\S\ref{sec:gammagamma}--\ref{sec:weibulltype}, we exploit the rate mixture construction \eqref{eq:ratio} and we propose new sub-asymptotic univariate tail models. In \S\ref{sec:gammagamma}, our proposed model is heavy-tailed with the GP distribution as a special case and we focus on it for spatial modeling in \S\S\ref{sec:spatial}--\ref{sec:application}, while in \S\ref{sec:weibulltype}, our proposed construction is a flexible model of Weibull type and has the GP distribution as a limiting boundary case.

\subsection{Gamma-gamma model}\label{sec:gammagamma}
Replacing the exponential distribution of $F_{Y\mid\Lambda}(\cdot)$ in \eqref{eq:GP.mixture} by a gamma distribution  yields the hierarchical gamma-gamma model, which may be written as
\begin{equation}
\label{eq:GammaGamma.mixture}
Y\mid \Lambda 
\sim {\rm Gamma}(\Lambda,\beta_1), \quad 
\Lambda
\sim {\rm Gamma}(\alpha,\beta_2), \quad \alpha,\beta_1,\beta_2>0; 
\end{equation}
i.e., $F_{Y\mid\Lambda=\lambda}(\cdot)$ is the ${\rm Gamma}(\lambda,\beta_1)$ distribution.
The model \eqref{eq:GammaGamma.mixture} simplifies to the GP distribution obtained in \eqref{eq:GP.mixture} when $\beta_1=1$. 
The distribution of $Y$ corresponds to a rescaled $F_{\nu_1,\nu_2}$ distribution with degrees of freedom $\nu_1=2\beta_1$ and $\nu_2=2\beta_2$, and scaling factor $\alpha\beta_1/\beta_2$, such that 
$Y\stackrel{D}{=} (\alpha\beta_1/\beta_2)Z$, with $Z\sim F_{2\beta_1,2\beta_2}$. Its density is
%
\begin{equation}
\label{eq:densityGammaGamma}
f(y)= \alpha^{-\beta_1}{\Gamma(\beta_1+\beta_2)\over\Gamma(\beta_1)\Gamma(\beta_2)}\left(1+{y\over\alpha}\right)^{-(\beta_1+\beta_2)} y^{\beta_1-1},\qquad y>0.
\end{equation}
The $r$-th moment of $Y$ is finite whenever $\beta_2>r$, and is given as
\begin{eqnarray}
\mathbb E(Y^r)
={{\alpha^r\,\Gamma(\beta_1+\beta_2)\,\Gamma(\beta_1+r)\,\Gamma(\beta_2-r)}\over{\Gamma(\beta_2)\,\Gamma(\beta_1)\,\Gamma(\beta_1+\beta_2+2)}},\qquad \beta_2>r.\nonumber 
\end{eqnarray}
From \eqref{eq:breiman}, or directly from \eqref{eq:densityGammaGamma}, we deduce that the gamma-gamma model has a heavy power-law tail. The tail index of the limiting GP distribution \eqref{eq:GP} is equal to $\xi=1/\beta_2$, and hence $\beta_2$ determines the tail decay rate of the distribution function $F$ of $Y$. Further details are provided in the Supplementary Material.

\subsection{Model extension with Weibull-type tail behavior}
\label{sec:weibulltype}
The gamma-gamma model \eqref{eq:GammaGamma.mixture} yields heavy tails (i.e., with a positive tail index, $\xi>0$) and thus has a relatively slow power-law tail decay. For data with a light upper tail (i.e., with a tail index equal to zero, $\xi=0$), we now discuss a flexible model extension based on the hierarchical construction \eqref{eq:ratio}, which provides a faster tail decay than the gamma-gamma model, while keeping the heavy-tailed GP distribution on the boundary of the parameter space. Specifically, we propose the following hierarchical model:
\begin{equation}
\label{eq:gammagig}
\begin{array}{rl}
Y^{1/k}\mid \Lambda &\sim {\rm Gamma}(\Lambda,\beta_1), \quad k,\beta_1>0,\\
\Lambda&\sim {\rm GIG}(\alpha/2,b,\beta_2), \quad (\alpha,b,\beta_2)\in D_{\mathrm{GIG}},
\end{array}
\end{equation}
where the latent rate parameter $\Lambda$ is assumed to follow the generalized inverse Gaussian (GIG) distribution with parameters $\alpha/2$, $b$ and $\beta_2$, and where $D_{\mathrm{GIG}}$ denotes its parameter space. More precisely, the GIG$(a,b,\beta)$ density is 
$g(y)=(a/b)^{\beta/2}\{2 K_\beta(\sqrt{ab})\}^{-1}y^{\beta-1}\exp\{-(ay+b/y)/2\}$, $a\geq 0$, $b\geq 0$, $y>0$, 
with parameter constraints on $\beta$ given by $-\infty<\beta<\infty$ if $a,b>0$, by $\beta>0$ if $b=0$ and  $a>0$, and by $\beta<0$ if $a=0$ and $b\geq 0$, and where $K_\beta$ denotes the modified Bessel function of second kind with parameter $\beta$. The GIG distribution has an exponentially decaying tail (i.e., Weibull-type tail with Weibull index one).



Model \eqref{eq:gammagig} can also be represented as the ratio $Y=(\tilde{Y}/\Lambda)^k$ with $\tilde{Y}\sim {\rm Gamma}(1,\beta_1)$ independent of $\Lambda\sim  {\rm GIG}(\alpha/2,b,\beta_2)$. This model generalizes the gamma-gamma construction in \eqref{eq:GammaGamma.mixture}, which is on the boundary of the parameter space with $b=0$, $k=1$ and $\beta_2>0$. Hence, the model captures a wide range of tail behaviors, from very light tails to relatively heavy tails. Specifically, when $b>0$, the random variables $\tilde{Y}^k$ and $1/\Lambda^k$ have Weibull-type tails, with Weibull indexes both equal to $1/k$.
From \eqref{eq:weibull}, we deduce that $Y$ has Weibull index $\eta_Y=(1/k^2)/(2/k)=1/(2k)>0$. 
Thus, when $b>0$, this model can capture any Weibull tail with any Weibull index, while when $b=0$, it can capture any power-law tail with any positive tail index $k/\beta_2>0$, thanks to  Breiman's Lemma \eqref{eq:breiman}. 


\section{A Bayesian spatial gamma-gamma model}
\label{sec:spatial}
%

\subsection{Bayesian hierarchical modeling framework} \label{sec:bayeshier}  

Accounting for spatial dependence is important for a variety of reasons, even if the precise estimation of the extremal dependence structure is of secondary importance. First, this allows to borrow strength across locations to reduce the uncertainty and improve the estimation of marginal distributions and of high quantiles. Second, a proper spatial model is needed whenever prediction at unobserved locations is required.

Let $Y(\bs)$, $\bs\in\mathcal S\subset\Real^2$, be the spatial process of interest, and assume that we observe it at finite set of locations $\bm s_1,\ldots,\bm {s}_d\in \mathcal S$. There are different approaches to model the dependence structure of $\bm Y=(Y_1,\ldots,Y_d)^T$, where $Y_j=Y(\bm s_j)\sim F_{Y_j}$. One possibility is to directly bind together the marginals of $Y_1,\ldots,Y_d$ through a \emph{copula} model (i.e., a multivariate distribution with standard uniform margins) without assuming any hierarchical structure; we call this the \emph{copula approach}. 
Let $C_{\bm Y}$ denote the underlying copula of the data ${\bm Y}$, which is unique if $\bm Y$ has a continuous distribution. Then $\bm Y$ has distribution function $F(y_1,\ldots,y_d)=C_{\bm Y}\{F_{Y_{1}}(y_1),\ldots,F_{Y_{d}}(y_d)\}$ and density $f(y_1,\ldots,y_d)=c_{\bm Y}\{F_{Y_{1}}(y_1),\ldots,F_{Y_{d}}(y_d)\}\prod_{j=1}^{d}f_{Y_j}(y_j),$ where $c_{\bm Y}$ is the copula density and $f_{Y_1},\ldots,f_{Y_d}$ are the marginal densities.
To model threshold exceedances with respect to a fixed threshold vector $\bm u =(u_1,\ldots,u_d)^T$, it is common to censor observations $Y_j$ falling below a corresponding threshold $u_j$. In the copula approach, the likelihood contribution of an observation $\bm y=(y_1,\ldots,y_d)^T$ such that $y_{j}\geq u_j$, $j=1,\ldots,j_0,$ and $y_{j}<u_j$, $j=j_0+1,\ldots,d,$ is
\begin{equation}\label{eq:censlik2}
{{\partial^{j_{0}}}\over{\partial{y_{1}}},\ldots,\partial{y_{j_{0}}}}F(\bm y) \biggl\vert_{ y_{j_{0}+1}= u_{j_{0}+1},\ldots, y_{d}= u_{d}}=\int_{-\infty}^{u_{j_{0}+1}} \ldots \int_{-\infty}^{u_d} f(y_1,\ldots,y_d) \text{d}{y_{j_{0}+1}}\ldots \text{d}{y_{d}}.
\end{equation}
When $j_0<d$, multivariate distribution functions must be calculated to evaluate \eqref{eq:censlik2}. For many copula models (e.g., Gaussian, Student's $t$), this requires expensive  multivariate numerical integrations. Then, the computational cost can become prohibitively high if  the dimension $d$ is large, and accuracy issues may arise. 

Here, we instead use the hierarchical approach, where the process $Y(\bm s)$ is assumed to be conditionally independent given a latent process $\Lambda(\bm s)$ with spatial dependence. Let $\Lambda_j=\Lambda(\bm s_j)$, $j=1,\ldots,d$, and $\bm \Lambda=(\Lambda_1,\ldots,\Lambda_d)^T$. We assume that, conditional on $\bm \Lambda$,  an observation $Y_j$ is independent of the other observations $Y_{j'}$, $j'\not=j$. The vector of latent variables $\bm \Lambda$, on the other hand, is specified through a copula $C_{\bm \Lambda}$. This \emph{hierarchical approach} with a \emph{latent copula}  separates the observed process $\bm Y$ from the latent variables $\bm\Lambda$. Note that in the Bayesian modeling literature, latent variables are often used to capture spatio-temporal patterns in marginal distributions (e.g., spatio-temporal trends), while here the latent process $\Lambda(\bm s)$ is part of the model formulation to ensure the required \emph{unconditional} marginal distributions (e.g., of gamma-gamma type as in \eqref{eq:GammaGamma.mixture}) and to capture spatial dependence. In particular, this implies that the latent variables involved in different replicates of the process $Y(\bm s)$ will also be different. Moreover, although the conditional independence assumption may be seen as a restriction, it still permits to capture a wide range of dependence structures (see \S\ref{sec:jointtail}), and it has significant computational benefits. By 
augmenting the data $\bm Y$ with the latent variables $\bm \Lambda$, the censored likelihood \eqref{eq:censlik2} can be formulated in terms of univariate censored terms for the conditionally independent components $Y_j\mid \Lambda_j$, while no censoring is required for $\bm \Lambda$. In \S\ref{sec:augment}, we exploit this latent variable approach for fully Bayesian inference  using Markov chain Monte Carlo (MCMC).

We now describe our proposed Bayesian spatial modeling framework in more detail. Let  $\bm{\Theta}=(\bm{\Theta}_{\bm Y}^T,\bm{\Theta}_{\bm{\Lambda}}^{\text{mar}^T},\bm{\Theta}_{\bm{\Lambda}}^{\text{cop}^T})^T$ be the  vector of unknown hyperparameters, where  $\bm{\Theta}_{\bm Y}$ controls the conditional distribution of observations,  and $\bm{\Theta}_{\bm{\Lambda}}=(\bm{\Theta}_{\bm{\Lambda}}^{\text{mar}^T},\bm \Theta_{\bm{\Lambda}}^{\text{dep}^T})^T
$ contains parameters for the latent process $\Lambda(\bm s)$, with $\bm{\Theta}_{\bm{\Lambda}}^{\text{mar}}$ controlling marginal distributions and   $\bm{\Theta}_{\bm{\Lambda}}^{\text{dep}}$ controlling the dependence structure. 
Our general spatial hierarchical construction is specified as 
\begin{eqnarray}
\label{eq:General.Hierarchical}
Y_j\mid {\bm \Lambda},\bm{\Theta}_{\bm Y} &\stackrel{\text{ind}}{\sim}& F_{Y} (\,\cdot\,;\Lambda_j,\bm {\Theta}_{\bm Y}),\qquad j=1,\ldots,d, \nonumber\\
\bm \Lambda\mid \bm \Theta_{\bm {\Lambda}} &\sim&   C_{\bm{\Lambda}}\left\{ F_{\Lambda_1}(\cdot;\bm \Theta_{\bm \Lambda}^{{\rm mar}}),\ldots,F_{\Lambda_d}(\cdot;\bm \Theta_{\bm \Lambda}^{{\rm mar}});\bm \Theta_{\bm\Lambda}^{{\rm dep}} \right\},\\ \nonumber
\bm{\Theta}&\sim & \pi(\bm{\Theta}),\nonumber
\end{eqnarray}
where $C_{\bm \Lambda}$ refers to the spatial copula  of $\bm \Lambda$,  $F_{\Lambda_j}(\cdot;\bm \Theta_{\bm \Lambda}^{{\rm mar}})$ denotes the marginal distribution of $\Lambda_j$, $j=1,\ldots,d$, and $\pi(\bm \Theta)$ is the prior distribution of the parameter vector $\bm \Theta$. 
The joint distribution of $\bm Y$, $\bm \Lambda$, and $\bm \Theta$ can be decomposed into conditional distributions as $\pi(\bm Y,\bm \Lambda, \bm \Theta_{\bm Y}, \bm \Theta_{\bm \Lambda})=\pi(\bm Y \mid \bm \Lambda,\bm \Theta_{\bm Y}) ~\pi(\bm \Lambda \mid\bm \Theta_{\bm \Lambda})~\pi(\bm \Theta)$, where $\pi(\cdot)$ denotes a generic (conditional) distribution. The joint posterior distribution $\pi\left(\bm \Lambda,\bm \Theta\mid \bm Y\right)$  of latent variables $\bm \Lambda$ and hyperparameters $\bm \Theta$ is then proportional to 
$\pi(\bm Y,\bm \Lambda, \bm \Theta)$, 
and the posterior distribution of hyperparameters $\bm \Theta$ is obtained by integrating out the latent parameters $\bm \Lambda$, i.e.,
\begin{eqnarray}
\label{eq:poterior.dist}
\pi \left(\bm \Theta\mid \bm Y\right)=\int \pi\left(\bm\Lambda, \bm \Theta\mid \bm Y\right)~ {\rm d} \bm{\Lambda}.
\end{eqnarray}
The dimension of the integration domain in \eqref{eq:poterior.dist} can be very high if there are a lot of latent variables.
We solve this issue in \S\ref{sec:bayesian} by implementing an MCMC algorithm, in which the latent variables $\bm \Lambda$ are imputed and updated at each iteration.


\subsection{A spatial gamma-gamma model} \label{sec:GammaGammaSpatialModel}
We now present a specific Bayesian hierarchical model of the form \eqref{eq:General.Hierarchical} that is based on 
the gamma-gamma construction \eqref{eq:GammaGamma.mixture} with a latent Gaussian copula process for the spatial dependence in $\bm \Lambda$. Here, $\bm \Theta_Y=\beta_1>0$, $\bm \Theta_{\bm \Lambda}^{{\rm dep}}$ contains the correlation parameters of the latent process $\Lambda(\bm s)$, and $\bm \Theta_{\bm \Lambda}^{{\rm mar}}=(\alpha,\beta_2)^T\in(0,\infty)^2$. We focus here on the isotropic exponential correlation function $\sigma(h)=\exp(-h/\rho),~h\geq 0,$ with range $\rho>0$, so that $\bm \Theta_{\bm \Lambda}^{{\rm dep}}=\rho$, although other correlation functions are also possible.
We write $\Phi$ for the univariate standard normal distribution and $\Phi_{\rho}$ for the zero mean and unit variance multivariate normal distribution associated to any collection of $d$ sites ${\bm s}_1,\ldots,{\bm s}_d$, and parametrized by the correlation matrix ${\bm \Sigma}(\rho)$ with entries ${\bm \Sigma}_{i_1i_2}=\sigma(\|{\bm s}_{i_1}-{\bm s}_{i_2}\|)=\exp(-\|{\bm s}_{i_2}-{\bm s}_{i_1}\|/\rho)$, $1\leq i_1,i_2\leq d$. The $\text{Gamma}(\alpha,\beta)$ distribution function is denoted by $\Gamma(\,\cdot\,; \alpha,\beta)$. Following \eqref{eq:General.Hierarchical}, we define the gamma-gamma hierarchical model with latent Gaussian copula as
\begin{eqnarray}
\label{eq:hierarchicalgammagamma}
Y_j\mid \bm \Lambda,\bm \Theta_{\bm Y} &\stackrel{\text{ind}}{\sim}& \Gamma(\cdot;\Lambda_j,\beta_1),\quad j=1,\ldots,d,\nonumber  \\ 
\bm \Lambda \mid \bm \Theta_{\bm \Lambda}&\sim &  \Phi_{\rho}[\Phi^{-1}\{\Gamma (\,\cdot\,;\alpha,\beta_2)\},\ldots,\Phi^{-1}\{\Gamma(\,\cdot\,;\alpha,\beta_2)\}],\\
\bm \Theta &\sim & \pi(\bm \Theta)=\pi(\alpha)\times\pi(\beta_1)\times\pi(\beta_2)\times\pi(\rho).\nonumber
\end{eqnarray}
While a Gaussian copula is specified in \eqref{eq:hierarchicalgammagamma}, other copula models with stronger tail dependence (e.g., the elliptically symmetric Student's $t$ copula with $\nu>0$ degrees of freedom and dispersion matrix ${\bm \Sigma}(\rho)$) are also possible. The implied extremal dependence structure is discussed in \S\ref{sec:jointtail}. Covariate information may be included in various ways into the model parameters. Here, because $\alpha$ describes the scale of the marginal distribution of the process $Y(\bm s)$, a  natural approach is to use a log-linear specification of the form
$
\log\alpha=\log \alpha_0+\alpha_1 x_1+\cdots+\alpha_p x_p$ for some known covariates $x_1,\ldots,x_p$. Similarly, we may add covariates information in the parameter $\beta_2$ as $\log\beta_2=\log \beta_{2;0}+\beta_{2;1} z_1+\cdots+\beta_{2;q} z_q$, where $z_1,\ldots,z_q$ are (potentially different) covariates. We next describe our choice of prior distributions for the hyperparameters of the gamma-gamma model \eqref{eq:hierarchicalgammagamma}.

\subsection{Prior distributions for hyperparameters}
\label{sec:pc}
Appropriate prior distributions in the model \eqref{eq:hierarchicalgammagamma} need to be designed for the components of the hyperparameter vector $\bm \Theta$, and special care is required for the shape parameters $\beta_1$ and $\beta_2$, which represent the ``distance'' to the GP model and the upper tail decay rate, respectively.
A possible choice is to select an informative prior distribution $\pi(\beta_1)$ for $\beta_1>0$ that shrinks our mixture model \eqref{eq:GammaGamma.mixture} towards the GP with  $\beta_1=1$, believed to be valid in the limit. 
This can be achieved through the concept of penalized-complexity (PC) priors \citep{Simpson.etal:2017}. PC priors assume  a constant-rate exponential distribution for the square root of the Kullback-Leibler divergence with respect to a simpler reference model. 
Let $\gamma(\cdot;\lambda,\beta_1)$ be the ${\rm Gamma}(\lambda,\beta_1)$ density and $\gamma(\cdot;\lambda,1)$ be the ${\rm Exp}(\lambda)$ density. The (asymmetric) Kullback-Leibler divergence of $\gamma(\cdot;\lambda,\beta_1)$ with respect to $\gamma(\cdot;\lambda,1)$ is
\begin{align}
{\rm KLD}\left\{\gamma(\cdot; \lambda,\beta_1)\| \gamma(\cdot;\lambda,1)\right\} &= \int_{0}^{\infty}\log\left\{{{\gamma(y;\lambda,\beta_1)}\over{\gamma(y;\lambda,1)}}\right\}\gamma(y;\lambda,\beta_1){\rm d}y, \nonumber \\  &= ({\beta_1}-1)\psi({\beta_1}) - \log\{\Gamma({\beta_1})\},\label{eq:KLD}
\end{align}
{
	where $\psi({\beta_1})={\rm d}\log\{\Gamma({\beta_1})\}/{\rm d}{\beta_1}$ denotes the polygamma function of order $0$, also known as the digamma function. From \eqref{eq:KLD}, the derivative of the Kullback-Leibler divergence is ${{\rm d}\over {\rm d} {\beta_1}}  {\rm KLD}\{\gamma(\cdot; \lambda,\beta_1)\| \gamma(\cdot;\lambda,1)\} = ({\beta_1}-1)\psi'({\beta_1})$ where $\psi'({\beta_1})={\rm d}\psi({\beta_1})/{\rm d}{\beta_1}$ is the polygamma function of order $1$. Writing $\ell({\beta_1})=\sqrt{ 2 {\rm KLD}\left\{\gamma(\cdot; \lambda,\beta_1)\| \gamma(\cdot;\lambda,1)\right\}}$ we deduce that the corresponding PC prior is a mixture of two densities defined over $0<\beta_1<1$ and $\beta_1>1$, i.e.,
	\begin{align}
	\pi(\beta_1)&={\kappa_1\over 2}\exp\{-\kappa_1\, \ell({\beta_1})\}\left|{{\rm d}\over {\rm d}\beta_1}\ell({\beta_1})\right|\label{eq:PCprior}\\
	&={\kappa_1\over 2}\exp\{-\kappa_1 \sqrt{2({\beta_1}-1)\psi({\beta_1}) - 2\log\{\Gamma({\beta_1})\}}\}\left|{({\beta_1}-1)\psi'({\beta_1})\over \sqrt{2({\beta_1}-1)\psi({\beta_1}) - 2\log\{\Gamma({\beta_1})\}}}\right|,\nonumber
	\end{align}
	for $\beta_1>0$, and where $\kappa_1>0$ is a predetermined  penalty rate. The PC prior \eqref{eq:PCprior} is displayed in Figure~\ref{fig:PCprior} for $\kappa_1=1,2,3$. As expected, the mass is concentrated near $\beta_1=1$.
	\begin{figure}
		\centering
		\includegraphics[width=0.8\linewidth]{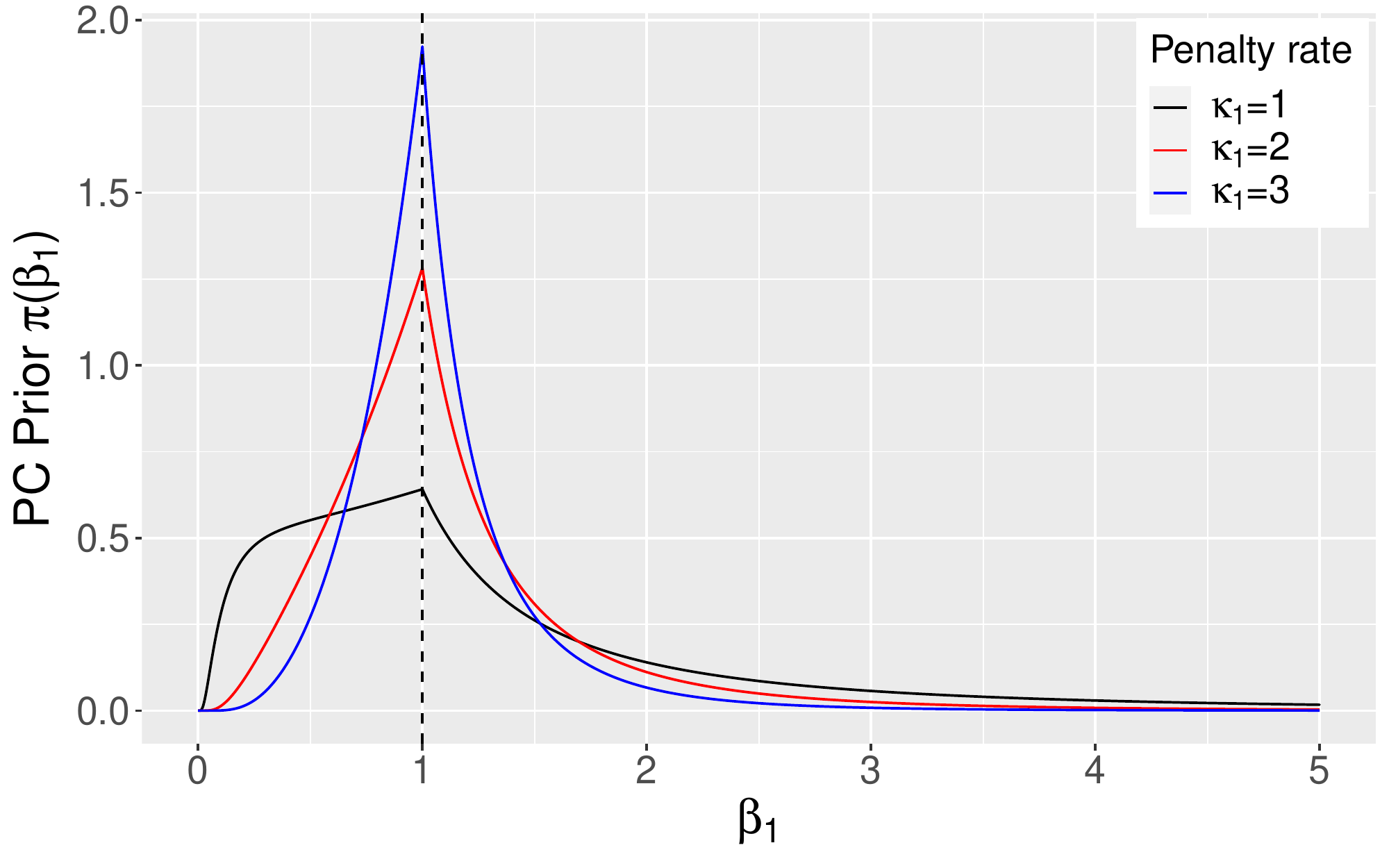}
		\caption{PC prior $\pi(\beta_1)$, $\beta_1>0$, as derived in \eqref{eq:PCprior}, for penalty rates $\kappa_1=1,2,3$ (black, red and blue curves, respectively).}
		\label{fig:PCprior}
	\end{figure}
	
	The prior distribution for $\beta_2$ is more conveniently constructed through the reparametrization given by the tail index $\xi=1/\beta_2$. It makes sense to prevent very heavy tails by shrinking $\xi$ towards zero (i.e., $\beta_2$ towards infinity), which corresponds to an exponential GP distribution in \eqref{eq:GP}. \citet{opitz2018inla} derived the PC-prior for $\xi$, which may be written as
	\begin{equation}
	\label{eq:PCprior2}
	\pi(\xi)=\sqrt{2}\kappa_2\exp\left\{-\sqrt{2}\kappa_2\,\xi(1-\xi)^{-1/2}\right\}(1-\xi/2)(1-\xi)^{-3/2}, \quad 0<\xi<1,
	\end{equation}
	where $\kappa_2$ is the penalty rate. As \eqref{eq:PCprior2} is compactly supported over the interval $(0,1)$, it prevents infinite-mean models. 
	A change of variables establishes that the PC prior for $\beta_2$ is
	\begin{equation}
	\label{eq:PCprior3}
	\pi(\beta_2)=\sqrt{2}\kappa_2\exp\left[-\sqrt{2}\kappa_2\{\beta_2(\beta_2-1)\}^{-1/2}\right](\beta_2-1/2)\{\beta_2(\beta_2-1)\}^{-3/2}, \quad \beta_2>1.
	\end{equation}
	Both PC priors \eqref{eq:PCprior2} and \eqref{eq:PCprior3} are illustrated in Figure~\ref{fig:PCprior2} for $\kappa_2=1,2,3$.
	
 We specify vague priors for  the other hyperparameters. More explicitly, we choose a gamma distribution with mean $1$ and variance $100$  for the correlation range $\rho$  and a Gaussian distribution with mean $0$ and variance $100$ for the covariate parameters, such as $\log \alpha_0,\alpha_1,\ldots,\alpha_p$ and $\log \beta_{2;0},\beta_{2;1},\ldots,\beta_{2;q}$. 
	\begin{figure}
		\centering
		\includegraphics[width=\linewidth]{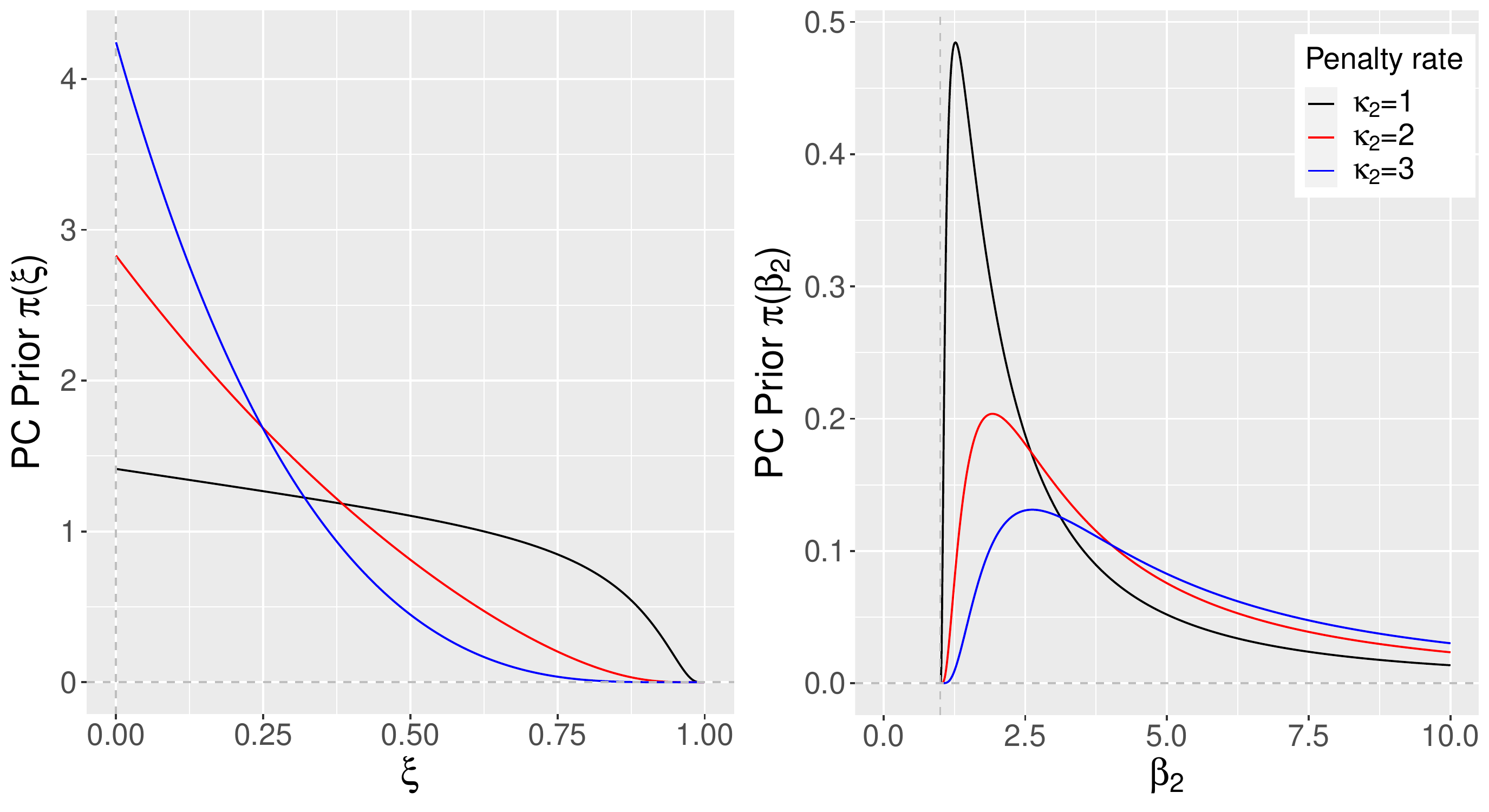}
		\caption{PC priors $\pi(\xi)$, $0<\xi<1$ (left) and $\pi(\beta_2)$, $\beta_2>1$ (right) as derived in \eqref{eq:PCprior2} and \eqref{eq:PCprior3}, respectively, for penalty rates $\kappa_2=1,2,3$ (black, red and blue curves, respectively).}
		\label{fig:PCprior2}
	\end{figure}
	

\subsection{Joint tail behavior}\label{sec:jointtail}
The upper-tail dependence in the hierarchical model with latent copula \eqref{eq:General.Hierarchical} is determined by the interplay between the lower joint tail  of $\bm \Lambda=(\Lambda_1,\ldots,\Lambda_d)^T$ and the  univariate upper tail  of the distribution $F_{Y\mid\Lambda=1}(\cdot)$ of the independent random variables $\tilde{Y}_j$, $j=1,\ldots,d$, stemming from \eqref{eq:ratio}.

We here provide more details for the case where $1/\Lambda(\bm s)$ has regularly varying distribution with positive tail index $\xi$, and $\tilde{Y}$ is lighter-tailed such that $\mathbb{E}(\tilde{Y}^{1/\xi+\varepsilon})<\infty$ for some $\varepsilon>0$, which includes the gamma-gamma model \eqref{eq:hierarchicalgammagamma}. If, in addition, the multivariate distribution $F_{1/\bm\Lambda}$ of $1/\bm \Lambda$ is regularly varying at infinity \citep{Resnick.1987}, we have
\begin{equation*}
{{1-F_{1/\bm \Lambda}(t\bm y)}\over{1-F_{1/\bm \Lambda}(t\bm 1)}}\rightarrow V_{1/\bm{\Lambda}}(\bm y), \quad \bm y>\bm 0, \quad t\rightarrow\infty,
\end{equation*}
where $\bm 1=(1,\ldots,1)^T\in \mathbb R^d$ and $V_{1/\bm{\Lambda}}(\bm y)$ is some positive limit function. Theorem 3 of \citet{Fougeres.Mercadier.2012} then implies multivariate regular variation of $F_{\bm Y}$, i.e.,		
\begin{align}
\label{eq:Jointtaillimit}
{{1-F_{\bm Y}(t\bm y)}\over{1-F_{\bm Y}(t \bm 1)}}\rightarrow V_{\bm Y}(\bm y)= \int_{0}^{\infty}\ldots\int_{0}^{\infty} V_{1/\bm{\Lambda}}(\bm y/\bm x) \prod_{j=1}^d F_{Y\mid \Lambda=1}(\mathrm{d} x_j), \qquad\bm y>\bm 0, \quad t\rightarrow\infty. 
\end{align}
The functions $V_{1/\bm{\Lambda}}$ and  $V_{\bm Y}$ are homogeneous of order $-1/\xi$, i.e., $V_{1/\bm{\Lambda}}(t\bm y)=t^{-1/\xi}V_{1/\bm{\Lambda}}(\bm y)$ and $V_{\bm Y}(t\bm y)=t^{-1/\xi}V_{\bm Y}	(\bm y)$  for positive values of $t$ and $\bm y$. }
Equation \eqref{eq:Jointtaillimit} fully characterizes the extremal dependence structure of the process $Y(\bm s)$ resulting from the construction \eqref{eq:General.Hierarchical} in the heavy-tailed case. Let $Y_1\sim F_{Y_1}$ and $Y_2\sim F_{Y_2}$, then a summary of the extremal dependence strength is the coefficient $\chi=\lim_{u\to 1} \chi(u)$, with $\chi(u)={\rm Pr}\{Y_1>F_{Y_1}^{-1}(u)\mid Y_2>F_{Y_2}^{-1}(u)\}$. It can be shown that $\chi=2-V_{\bm Y}[\{V_{\bm Y}(\infty,1)\}^{\xi},\{V_{\bm Y}(1,\infty)\}^\xi]$, where $\bm Y=(Y_1,Y_2)^T$. The pair of variables $\bm Y$ is called asymptotically independent if $\chi=0$ and asymptotically dependent if $\chi>0$. The case of asymptotic independence corresponds to a  $V_{\bm Y}$ function  that is a sum of separate terms for the components, i.e., $V_{\bm Y}(y_1,y_2)=c(y_1^{-1/\xi}+y_2^{-1/\xi})$ with a constant $c> 0$.  From \eqref{eq:Jointtaillimit}, it follows that $\bm Y$ is asymptotically independent if and only if $1/\bm \Lambda=(1/\Lambda_1,1/\Lambda_2)^T$ is asymptotically independent. Therefore the gamma-gamma model \eqref{eq:hierarchicalgammagamma} with latent Gaussian copula is asymptotically independent. When the Student's $t$ copula with $\nu>0$ degrees of freedom is used instead, the process $Y(\bm s)$ becomes asymptotically dependent (despite the conditional independence assumption at the data level).

\begin{figure}
		\centering
		\includegraphics[width=\linewidth]{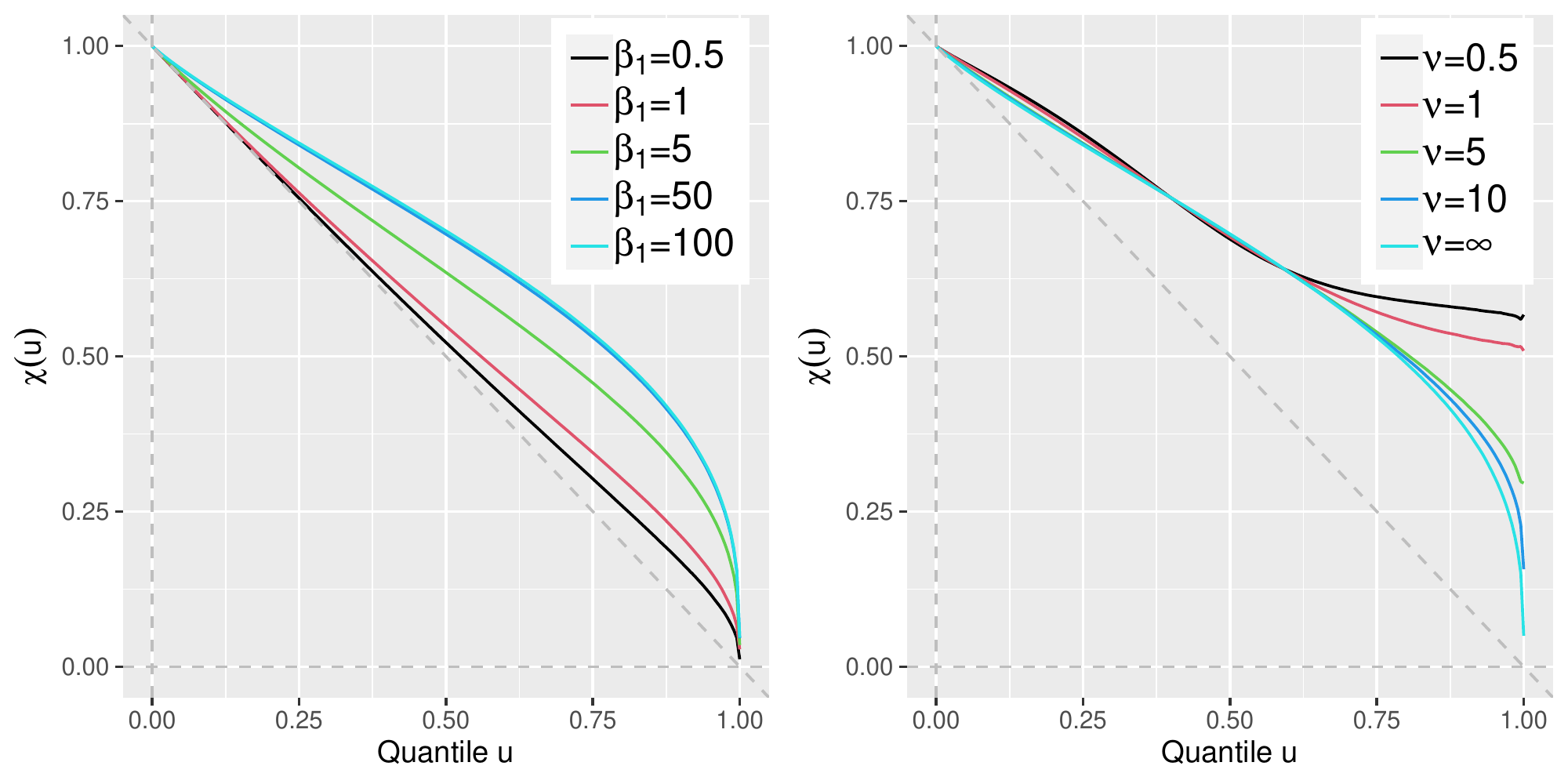}
		\caption{Plot of $\chi(u)$ against the quantile $u\in(0,1)$ for the gamma-gamma model  \eqref{eq:hierarchicalgammagamma} at spatial distance $0.5$ with latent Gaussian copula (left) and parameters $\alpha=1,\beta_1=0.5,1,5,50,100,\beta_2=2.5$ and $\rho=1$, and Student's $t$ copula (right) with degrees of freedom $\nu=0.5,1,5,10,\infty$ (Gaussian) and $\alpha=1$,
  $\beta_1=50$, $\beta_2=2.5$ and $\rho=1$. The dashed diagonal gray lines correspond to exact independence.}
		\label{fig:ChiMeasure}
	\end{figure}

To illustrate the dependence strength of the gamma-gamma model \eqref{eq:hierarchicalgammagamma}, we compute $\chi(u)$, with $u\in(0,1)$, by simulation for different parameter values. The left panel of Figure \ref{fig:ChiMeasure} shows $\chi(u)$ obtained at spatial distance $0.5$ using a Gaussian copula with correlation range $\rho=1$, and the other hyperparameters set to $\alpha=1$, $\beta_1=0.5,1,5,50,100$, $\beta_2=2.5$ (i.e., $\xi=0.4$). The right panel of Figure \ref{fig:ChiMeasure} shows $\chi(u)$ obtained using a Student's $t$ latent copula with range $\rho=1$ and degrees of freedom $\nu=0.5,1,5,10,\infty$ (Gaussian), and the other hyperparameters set to $\alpha=1$, $\beta_1=50$, $\beta_2=2.5$. These plots demonstrate that our hierarchical modeling approach can capture various joint tail decay rates and extremal dependence structures.

\section{Simulation-based Bayesian inference}
\label{sec:bayesian}

\subsection{General strategy}
\label{sec:GE}

We use Markov chain Monte Carlo (MCMC) sampling to generate a representative posterior sample of the hyperparameter vector $\bm \Theta$ and the latent variables $\bm\Lambda$ involved in the hierarchical model \eqref{eq:General.Hierarchical}, conditional on observed data.  
As we fit the model to threshold exceedances, we first describe the censored likelihood mechanism with latent variables in \S\ref{sec:augment}, focusing on the gamma-gamma model \eqref{eq:hierarchicalgammagamma}. Then in \S\ref{sec:mcmc}  we develop an efficient MCMC sampler by using the Metropolis-adjusted Langevin algorithm (MALA) for generating MCMC block proposals that ensure a relatively fast exploration of the high-dimensional parameter space of $\bm{\Lambda}$. Also, we propose an adaptive algorithm \citep[see][for an overview]{Andrieu.Thoms:2008} to tune the calibration parameters of the MALA and random walk proposals for an appropriate convergence rate.

	\subsection{Censored likelihood with latent variables}
	\label{sec:augment}
	We suppose that observed data $Y_i(\bm s_j)$, $i=1,\ldots, n$, $j=1,\ldots,d$, are composed of $n$ independent time replicates of the $d$ components of a random vector  $\bm Y=\{Y(\bm s_1),\ldots,Y(\bm s_d)\}^T$ indexed by locations $\bm s_1,\ldots,\bm s_d$. 
	We discuss here inference for the spatial hierarchical gamma-gamma model with latent Gaussian copula in \eqref{eq:hierarchicalgammagamma}, although little would change for other hierarchical models of the form \eqref{eq:General.Hierarchical}.  We write $y_{ij}=Y_i(\bm s_j),\lambda_{ij}=\Lambda_i(\bm s_j)$, $i=1,\ldots,n$, $j=1,\ldots,d$, and use the symbols $\phi$ and $\phi_{\rho}$ for the univariate and multivariate Gaussian densities corresponding to $\Phi$ and $\Phi_{\rho}$, respectively. Given a data vector $\bm y_i=(y_{i1},\ldots,y_{id})^T$ and a fixed threshold vector $\bm u_i=(u_{i1},\ldots,u_{id})^T\in[0,\infty)^d$, we introduce the exceedance indicator vector $\bm e_i=(e_{i1},\ldots,e_{id})^T$ with $e_{ij}=1$ if $y_{ij}\geq u_{ij}$ and $e_{ij}=0$ otherwise. If $u_{ij}=0$, no censoring is applied to the value $y_{ij}$,  on the other hand, if  $u_{ij}=\infty$ then the observation $y_{ij}$ is treated as fully censored. This may be used to handle missing data and prediction at unobserved locations.
	We now give the augmented censored  likelihood contribution of $\bm y_i$ where we consider both $\bm \Theta$ and $\bm\lambda_i=(\lambda_{i1},\ldots,\lambda_{id})^T$ as parameters. The density of observations $(y_{ij},e_{ij})$ conditional on $\lambda_{ij}$ is 
	$f_c(y_{ij},e_{ij};\lambda_{ij},\beta_1)=\Gamma(u_{ij};\lambda_{ij},\beta_1)$ if $e_{ij}=0$ and $f_c(y_{ij},e_{ij};\lambda_{ij},\beta_1)=\gamma(y_{ij};\lambda_{ij},\beta_1)$ if $e_{ij}=1$, 
	where $\gamma(\,\cdot\,;\lambda_{ij},\beta_1)$ is the gamma density with parameters $\lambda_{ij}$ and $\beta_1$. 
	The augmented censored likelihood contribution for the data vector $(\bm y_i^T,\bm e_i^T)^T$ is thus
	\begin{align}
	L(\bm \Theta,\bm \lambda_i;\bm y_i,\bm e_i)=&\prod_{j=1}^{d} f_c(y_{ij},e_{ij};\lambda_{ij} ,\beta_1)\\
	&\times \phi_{\rho}[\Phi^{-1}\{\Gamma(\lambda_{i1};\alpha,\beta_2)\},\ldots,\Phi^{-1}\{\Gamma(\lambda_{id};\alpha,\beta_2)\}] \times \prod_{j=1}^d{{\gamma(\lambda_{ij};\alpha,\beta_2)}\over{\phi[\Phi^{-1}\{\Gamma(\lambda_{ij};\alpha,\beta_2)\}]}} \nonumber,
	\end{align}
	where the first line refers to the observation model and the second line to the latent model. The overall augmented censored likelihood is
	\begin{equation}\label{eq:censlik}
	L_n(\bm \Theta,\bm \lambda ; \bm y,\bm e)=\prod_{i=1}^n L(\bm \Theta, \bm\lambda_i; \bm y_i,\bm e_i),
	\end{equation}
	where $\bm \lambda=(\bm \lambda_1^T,\ldots,\bm \lambda_n^T)^T$, $\bm y=(\bm y_1^T,\ldots,\bm y_n^T)^T$, and $\bm e=(\bm e_1^T,\ldots,\bm e_n^T)^T
$.	Notice that thanks to data augmentation and to the conditional independence assumption, only univariate censoring is required, hence facilitating computations.

	\subsection{Metropolis--Hastings MCMC algorithm with adaptive MALA and random walk proposals}
	\label{sec:mcmc}
	
	
We implement a Metropolis--Hastings MCMC algorithm 
	to sample from the posterior distribution of hyperparameters $\bm \Theta$ and latent variables $\bm \lambda=(\bm \lambda_1^T,\ldots, \bm \lambda_n^T)^T$. More precisely, we update the parameters $\bm \Theta$ and $\bm \lambda$ in two separate blocks for a predetermined number of iterations, in order to construct a Markov chain whose stationary distribution is the posterior distribution of interest. 
	To avoid invalid proposals or strong dependence between posterior samples of latent parameters or hyperparameters, we first apply the following reparametrization of the model: the latent parameters are log-transformed, i.e., $\tilde{\bm\lambda}=\log(\bm\lambda)$, while
	the hyperparameters of the gamma-gamma model~\eqref{eq:hierarchicalgammagamma} are reparametrized (internally) as
	\begin{equation}\label{eq:transf}
	\tilde{\alpha} =\log(\alpha\beta_1/\beta_2), \quad 
	\tilde{\beta_1} =\log(\alpha\beta_1^2/\beta_2), \quad 
	\tilde{\beta_2} =\log(1/\beta_2), \quad 
	\tilde{\rho} =\log(\rho).
	\end{equation}
The reverse transformation is $\alpha =\exp(2\tilde{\alpha})\exp(-\tilde{\beta_1})\exp(-\tilde{\beta_2})$, $\beta_1 =\exp(-\tilde{\alpha})\exp(\tilde{\beta_1})$, $\beta_2 =\exp(-\tilde{\beta_2})$, $\rho =\exp(\tilde{\rho})$. To use this modified internal reparametrization, we correct the target posterior density $\pi_{\text{post}}(\bm \Theta,\bm{\lambda}\mid \bm y, \bm e)\propto L_n(\bm \Theta,\bm \lambda; \bm y,\bm e) \pi(\bm \Theta)$ through the determinant of the Jacobian matrix of the transformation; its value is  $\exp(\tilde{\rho}+\tilde{\alpha}-2\tilde{\beta_2})$ for the hyperparameter transformation in \eqref{eq:transf}.

Our proposed MCMC algorithm consists of the following steps: we iteratively propose candidate values for the transformed hyperparameters $\bm{\tilde{\Theta}}$ and latent parameters $\bm{\tilde{\lambda}}$ from some proposal densities $q_1(\bm{\tilde{\Theta}}'\mid \bm{\tilde{\Theta}})$ and $q_2(\bm{\tilde{\lambda}}'\mid \bm{\tilde{\lambda}})$, respectively, and we accept these candidates with probability 
	\begin{equation*}
	\label{eq:accrate}
\min\left(1,{{L_n(\bm{\tilde{\Theta}}',\bm{\tilde{\lambda}}; \bm y,\bm e)\,\pi(\bm{\tilde{\Theta}}') \, q_1(\bm{\tilde{\Theta}}\mid \bm{\tilde{\Theta}}') }\over{L_n(\bm{\tilde{\Theta}},\bm{\tilde{\lambda}};\bm y,\bm e)\,\pi(\bm{\tilde{\Theta}}) \, q_1(\bm{\tilde{\Theta}}'\mid \bm{\tilde{\Theta}})}}\right),\qquad\min\left(1,{{L_n(\bm{\tilde{\Theta}},\bm{\tilde{\lambda}}'; \bm y,\bm e)\, q_2(\bm{\tilde{\lambda}}\mid\bm{\tilde{\lambda}}') }\over{L_n(\bm{\tilde{\Theta}},\bm{\tilde{\lambda}};\bm y,\bm e) \, q_2(\bm{\tilde{\lambda}}'\mid\bm{\tilde{\lambda}}) }}\right),
	\end{equation*}
	for hyperparameters and latent parameters, respectively.
	The number of parameters (latent variables and hyperparameters) to be explored by the Markov chain is equal to $N=nd+l$, with $l=|\bm{\tilde{\Theta}}|$. In particular,  it grows linearly with the sample size $n$ and dimension $d$. To handle the high dimensionality of the vector of latent variables, we propose using the Metropolis-adjusted Langevin algorithm (MALA), which exploits the gradient of the log-posterior density evaluated at the current parameter configuration to design an efficient multivariate Gaussian proposal density $q_2(\bm{\tilde{\lambda}}'\mid\bm{\tilde{\lambda}})$. Because the number of hyperparameters is moderate, we specify simple random walk proposals for $q_1(\bm{\tilde{\Theta}}'\mid\bm{\tilde{\Theta}})$.
Specifically, we propose candidate values $\bm{\tilde{\Theta}}'$ and $\bm{\tilde{\lambda}}'$ consecutively as follows:
	\begin{align*}
	\bm{\tilde{\Theta}}'\mid \bm{\tilde{\Theta}}&\sim \mathcal{N}(\bm{\tilde{\Theta}}, \tau_{\bm{\Theta}} I_l),  \\
	\bm{\tilde{\lambda}}'\mid \bm{\tilde{\Theta}},\bm{\tilde{\lambda}}&\sim \mathcal{N}(\bm{\tilde{\lambda}} + \tau_{{\bm \lambda}} \nabla_{\bm{\tilde{\lambda}}} \log \pi_{\text{post}}(\bm{\tilde{\Theta}},\bm{\tilde{\lambda}}\mid \bm y,\bm e), 2\tau_{{\bm \lambda}} I_{nd}), 
	\end{align*}
	where $I_l$ and $I_{nd}$ are the identity matrices of  dimensions $l\times l$ and $nd\times nd$, respectively, and $\tau_{\bm \Theta}>0$ and $\tau_{\bm \lambda}>0$ are step sizes controlling the variance of $q_1$ and $q_2$, respectively. In our proposed model, the gradient of the log-posterior density can be obtained in closed form, facilitating inference; see the details in the Supplementary Material. 
We use two burn-in phases in our MCMC algorithm. During the initial burn-in phase, we adapt  the tuning parameters $\tau_{\bm \Theta}$ and $\tau_{\bm \lambda}$ as follows. Let $\tau_{\text{cur}}$ denote the current value of either $\tau_{\bm \Theta}$ or $\tau_{\bm \lambda}$, $P_{\text{acc}}$ be the current acceptance probability calculated from the last $500$ iterations, and $P_{\text{tar}}$ be a target acceptance probability. Every $500$ iterations, we update $\tau_{\text{cur}}$ as $\tau_{\text{cur}}\mapsto\tau_{\text{new}}:=\exp\left\{({{P_{\text{acc}}-P_{\text{tar}}})/{\omega}}\right\}\tau_{\text{cur}}$, where  $\omega$ controls the rate of change. Here, we set $P_{\text{tar}}=0.57$ for $\tau_{\bm \lambda}$, which was found to be optimal for the MALA algorithm under independence assumptions \citep{roberts1998optimal}, and $P_{\text{tar}}=0.23$ for $\tau_{\bm \Theta}$, which usually works well for random walks. Moreover, we here fix $\omega$ to $0.4$. In the second burn-in phase, we use the same adaptive scheme only if the acceptance probability drops out of the intervals $[0.50,0.65]$ and $[0.15,0.30]$ for MALA and random walk proposals, respectively, i.e., if it has not stabilized yet during the initial burn-in phase. In all simulation experiments, we use a total of $1,500,000$ iterations, with $250,000$ iterations for the initial burn-in phase, and $500,000$ iterations for the second burn-in phase, whereas for the data application we doubled the length of all phases.
\section{Simulation study}
\label{sec:simulation}

%
%
\subsection{Simulation scenarios}
\label{sec:simulationSetting}
In this section, we study the performance of our MCMC sampler under diverse simulation scenarios. Data are simulated from the gamma-gamma model \eqref{eq:hierarchicalgammagamma} with latent copula model for $d\in\{50,100,200,400\}$ spatial locations sampled at random (i.e., uniformly) in the unit square $[0,1]^2$, with $n\in\{50,100,200,250\}$ temporal replicates, depending on the scenario. The latent process $\Lambda(\bm s)$ has a Gaussian or Student's $t$ copula with isotropic exponential correlation function $\sigma({\bm s}_i,{\bm s}_j)=\exp\{-(\|{\bm s}_i-{\bm s}_j\|/\rho)\},~\rho>0$. In some cases, we apply the censoring scheme presented in \S\ref{sec:augment}, using site-specific thresholds chosen as the empirical $75\%$-quantile. When performing spatial prediction, we set the censoring threshold to $+\infty$.

To be concise, we here only discuss the results for one scenario, and we report the results for all scenarios in the Supplementary Material. 
Specifically, we here consider $d=100$ spatial locations ($80$ used for fitting, and $20$ kept for spatial prediction), $n=100$ temporal replicates, and we use a latent Gaussian copula with range $\rho=1$. We assume that the marginal scale parameter depends on spatially-varying covariates in log-linear specification, such that $\alpha(\bm s)=\alpha_0\exp\bigl\{\sum_{k=1}^3\alpha_k z_k(\bm s)\bigr\}$, with covariates $z_1(\bm s)$ and $z_2(\bm s)$ corresponding to the $x$ and $y$ coordinates of the location $\bm s$, respectively, and $z_3(\bm s)$ being generated as a Gaussian random field model with mean $0$, variance $1$, and correlation $\sigma_{ij}=\exp(-\|\bm s_i-\bm s_j\|/2)$. This setting is similar to the setting considered in the application in \S\ref{sec:application}. The regression coefficients are set to $\alpha_0=\alpha_1=\alpha_2=\alpha_3=1$ and the other hyperparameters are fixed to $\beta_1=5$ and $\beta_2=5$  (i.e., with tail index $\xi=0.2$). 

We use the adaptive MCMC algorithm developed in \S\ref{sec:mcmc}. 
 We run two MCMC chains with different initial values in parallel to check the dependence on initial conditions,  and we then calculate MCMC outputs by combining these two chains.
\subsection{Results}
\label{sec:SimResult}
\begin{figure}[t!]
\centering
\includegraphics[width=\linewidth]{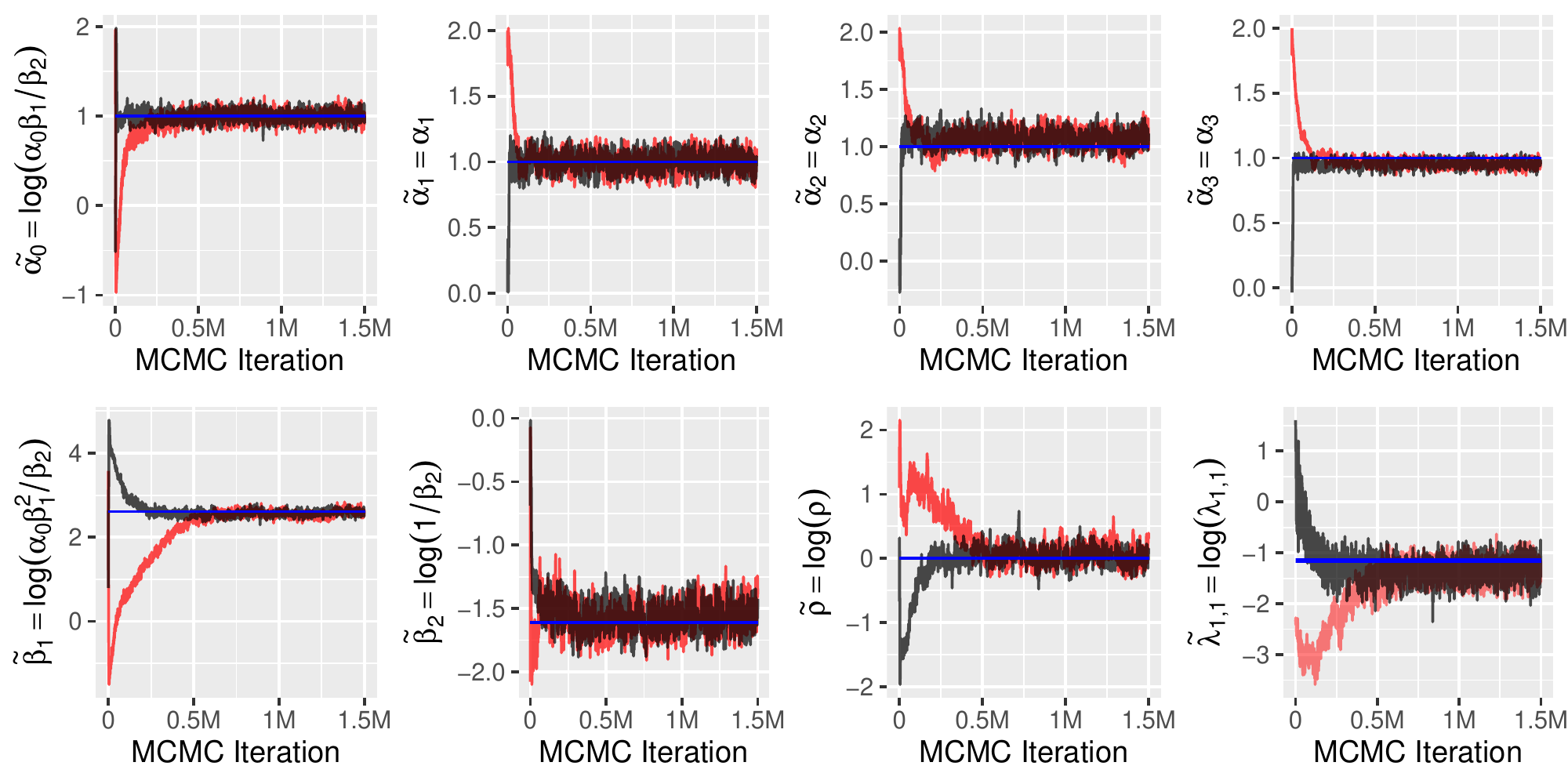}
\caption{Trace plots for the simulation scenario detailed in \S\ref{sec:simulationSetting}, where the hyperparameters are set to $\alpha_0=\alpha_1=\alpha_2=\alpha_3=1$, $\beta_1=5$, $\beta_2=5$ (i.e., $\xi=0.2$), and $\rho=1$. The rightmost plot in the second row  corresponds to  one chosen latent parameter, and the remaining $7$ plots are for all the hyperparameters. The red and black chains are two MCMC samples with different initial values, and the horizontal blue line represents the true values of the parameters. The total number of MCMC samples is $1,500,000$.}
\label{fig:TracePlotPrediction}
\end{figure}
Figure \ref{fig:TracePlotPrediction}  displays the trace plots for all the hyperparameters and one selected latent variable, for two MCMC chains with different initial values. The results show that there is good mixing and that convergence occurs very quickly for the regression parameters $\alpha_0$, $\alpha_1$, $\alpha_2$, and $\alpha_3$, and relatively quickly for the other hyperparameters or latent variables, despite the large number of latent variables. Essentially in this scenario, all Markov chains seem to have converged after about $350,000$ to $500,000$ iterations. For all parameters, the true value lies well within the corresponding posterior distribution, confirming the good performance of our algorithm. The effective sample size per minute is roughly between $10$ and $60$, see Table~1 in the Supplementary Material. The proposal variances, automatically tuned in our algorithm, converged to the values $\tau_{\bm \Theta}=1.2\times 10^{-4}$ for the hyperparameters (based on random walk proposals) and $\tau_{\bm \lambda}=5.4\times 10^{-4}$ for the latent variables (based on MALA proposals). The total run time for each Markov chain is almost $1.5$ days to achieve  $1.5$ million iterations.
\begin{figure}[t!]
\centering
\includegraphics[width=\linewidth]{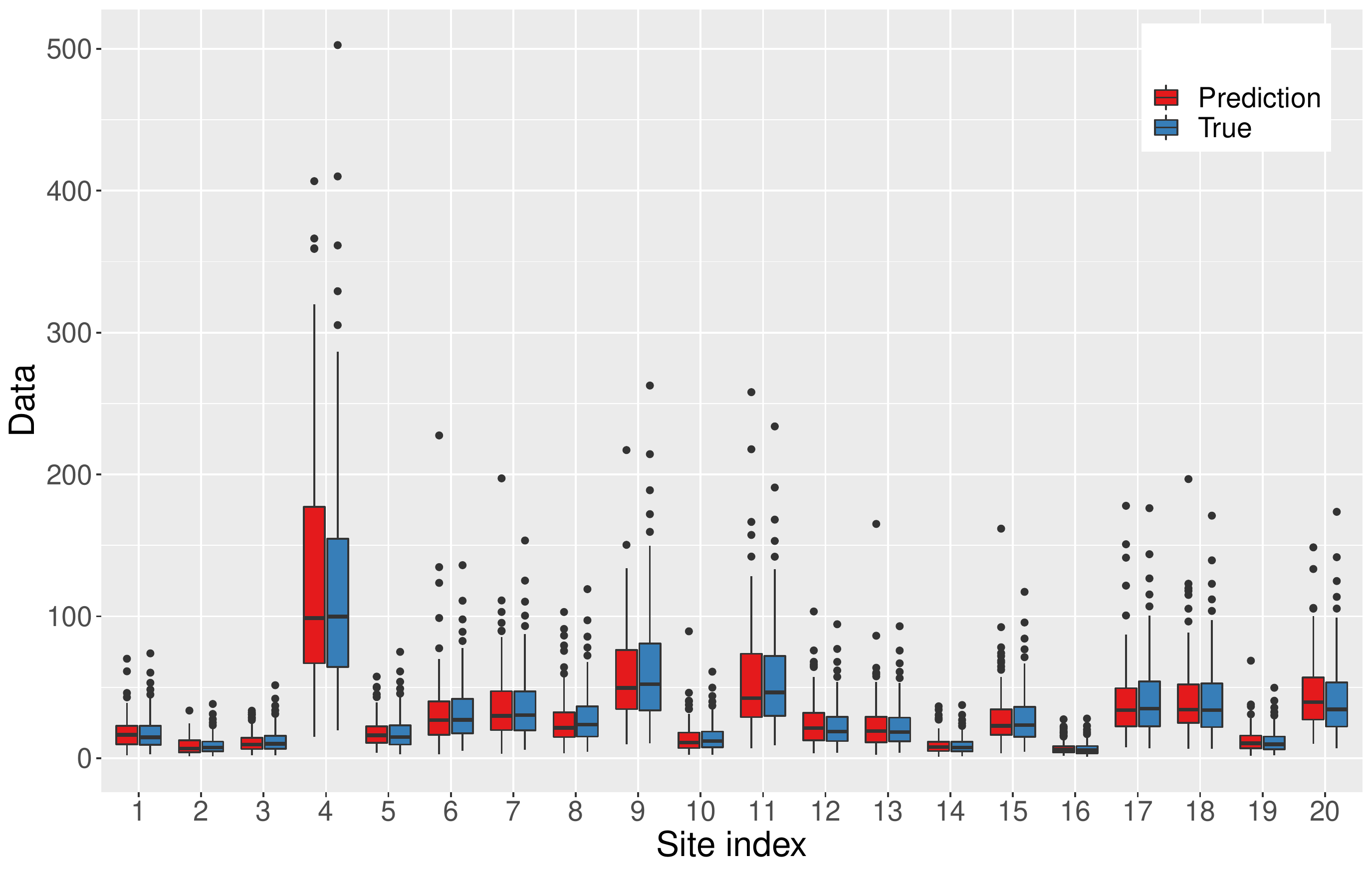}
\caption{Boxplots of the posterior predictive samples (red) and true observations (blue) at all the prediction sites, for the simulation scenario described in \S\ref{sec:simulationSetting}. The total number of MCMC samples is $1,500,000$, and the posterior samples are derived from the last $750,000$ MCMC samples, obtained after removing the initial $750,000$ burn-in samples.}\label{fig:PredictSites}
\end{figure}

 We then study the ability of our model to predict values at unobserved locations. To this end, we treat the data at $20$ randomly selected sites as missing, and we compute the posterior predictive distributions at these sites. 
Figure \ref{fig:PredictSites} compares boxplots of the true simulated values at the $20$ prediction sites to posterior predictive samples obtained from fitting our model. Clearly, the posterior predictive distributions at all prediction sites appropriately capture the natural variability in the true data. This suggests that our algorithm succeeds in performing spatial prediction. The great benefit of our Bayesian approach is that estimation based on censored data and spatio-temporal prediction are performed simultaneously.
\section{Application to precipitation extremes from Germany}
\label{sec:application}
\begin{figure}[t!]
 \centering
\includegraphics[width=\linewidth]{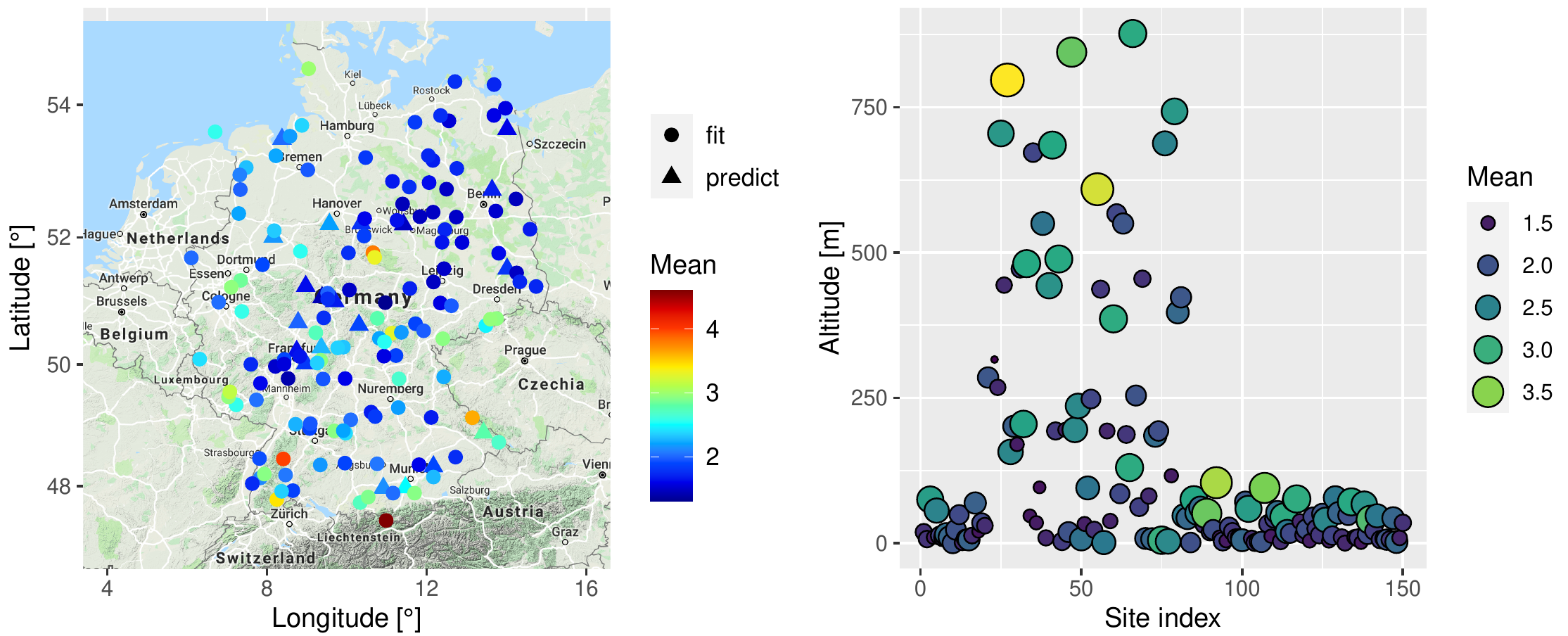}
\caption{Left: Mean precipitation [mm] at the study sites calculated over the days corresponding to the selected extreme events; see \S\ref{sec:extmocc}. Right: Mean precipitation [mm] plotted with respect to the station's altitude [m]. For visibility purposes, we do not show the station at the highest altitude ($\approx 3000$m) on the right panel.}
\label{fig:SpatialMeanPlot}
\end{figure}


\subsection{ Data description}
We now study precipitation intensities observed in  Germany, publicly available from the \href{https://www.ecad.eu/}{European Climate Assessment \& Dataset} project. 
The dataset reports daily precipitation amounts observed at more than $5000$ spatial locations during the period 1941 to 2018. We apply our hierarchical models to a subset of $d=150$ locations with some missing data
for the study period from 2009 to 2018. To avoid modeling complex seasonal non-stationarities, we consider the observations for the (full) months of September to December (i.e., for the extended autumn season), resulting in $n=1220$ temporal replicates. This time period was merely selected based on temporal stationarity diagnostics, although from a practical perspective it would also be interesting in future research to extend our stationary models in order to study seasonal patterns in precipitation intensities during other months, as well as to assess the flood risk all year round. The site-specific mean precipitation intensities reported in  Figure \ref{fig:SpatialMeanPlot} show a tendency towards higher values in regions with higher altitudes; see \citet{Murawski.etal:2016} for more details about spatial and temporal trends in precipitation over Germany. 
In a preliminary analysis of the tail behavior of precipitation intensities, we fit the  GP distribution to exceedances at each site separately with site-specific thresholds fixed at the $85\%$ empirical quantile of positive precipitation intensities. The maximum likelihood estimator, and  the moment-based estimator of \citet{dekkers1989moment} of the tail index,  both provide systematically positive tail index estimates. This finding suggests that precipitation intensities are heavy-tailed as expected, and we proceed by fitting the spatial gamma-gamma model \eqref{eq:hierarchicalgammagamma} to selected extreme precipitation events (see \S\ref{sec:gammaGammaAppl}). The selection of extreme events and the modeling of their occurrences are described in \S\ref{sec:extmocc}. 

%
%

 \subsection{Identifying and modeling extreme precipitation occurrences}
\label{sec:extmocc}
The precipitation intensities are zero or very small for most of the days in the observation period, and we first extract extreme events (i.e., specific days) used to fit our spatial hierarchical model. Let  $Y_t(\bm s_j)$ denote the precipitation intensities at time $t$ and site $\bm s_j$, $j=1,\ldots, d$. To select   extreme events, we consider the spatial average precipitation $S_t=d^{-1}\sum_{j=1}^{d}Y_t(\bm s_j)$, indexed by time $t$. We then define extreme events as days $t$ such that $S_{t}>\hat{G}^{-1}(0.85)$, where  $\hat{G}$ is the empirical cumulative distribution function of the sample of $S_{t}$ values. This  scheme 
extracts $181$ extreme events in total. 

$
$
Let $E_{a,t}$ denote the binary sequence of $0$ and $1$ values representing the occurrence indicators of extreme events for the $a$-th year, with $a\in\{1,\ldots,10\}$. That is, $E_{a,t}=1$ if the $t$-th day of the year, with $t\in\{244,\ldots,365\}$, was extreme in year $a$, and $E_{a,t}=0$ otherwise. 
To capture temporal dependence, we model this time series through a logistic regression with a random effect defined as a first-order autoregressive Gaussian process, i.e.,
\begin{align}
    \log\left\{{\pr(E_{a,t}=1)\over1-\pr(E_{a,t}=1)}\right\}&=\beta_{E,0}+ W_{a,t}, \label{eq:logreg}\\
    W_{a,1}&\iid \mathcal{N}(0,\sigma_\varepsilon^2/(1-\rho_E^2)), \quad a=1,\ldots, 10,\notag\\
    W_{a,t}\mid W_{a,t-1}&=\rho_E W_{a,t-1}+\varepsilon_{a,t},\quad t=244,\ldots,365, \notag\\ 
    \varepsilon_{a,t}&\stackrel{\text{iid}}{\sim}\mathcal{N}(0,\sigma_\varepsilon^2),\notag
\end{align}
with global intercept $\beta_{E,0}$, autoregression coefficient $\rho_E\in(-1,1)$, and marginal variance of $W_{a,t}$, $\sigma^2_E=\sigma_\varepsilon^2/(1-\rho_E^2)>0$. 
We have explored a number of alternative models including an additional seasonal trend component in the regression equation \eqref{eq:logreg}, but we could not detect any significant improvement with respect to model \eqref{eq:logreg}. 

\begin{figure}
    \centering
    \includegraphics[width=\linewidth]{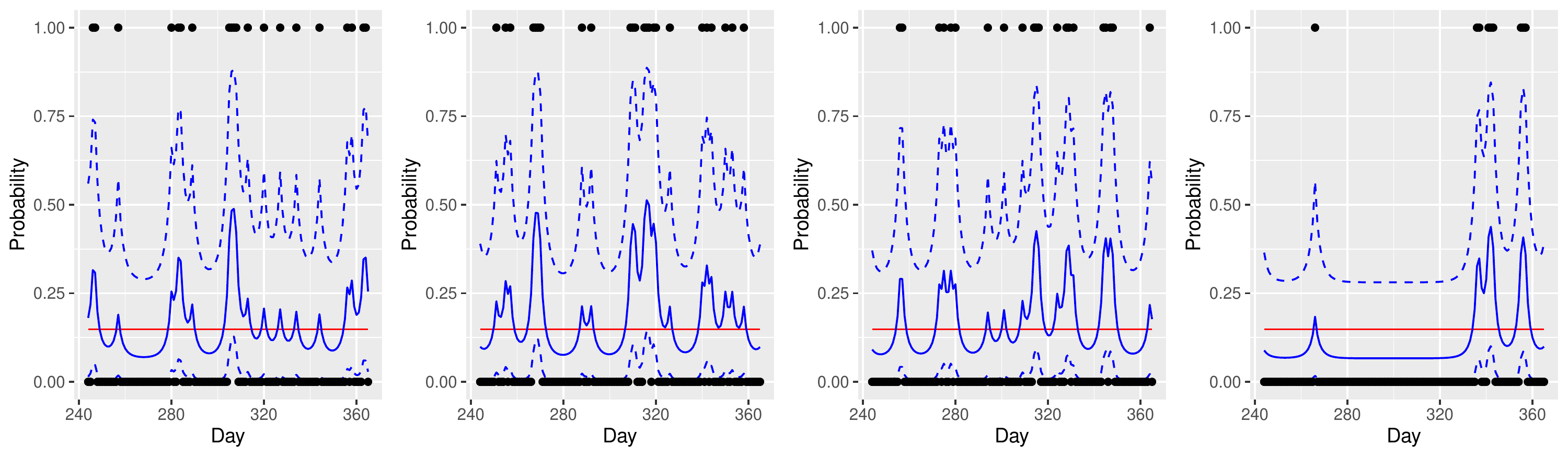}
    \caption{Posterior estimates (fitted probability values) of observing an extreme event at a given day according to model \eqref{eq:logreg} for the years $1,2,9,10$ (from left to right). Black $\bullet$ symbols indicate the observations $E_{a,t}$. Posterior means and  $95\%$ credible intervals of $\hat{\pr}(E_{a,t}=1)$ are reported by continuous and dashed blue curves, respectively. The red line indicates the global empirical exceedance probability equal to $0.15$.}
    \label{fig:pfitted}
\end{figure}

We now present the estimation results for model \eqref{eq:logreg}. While there would be no notable obstacles for MCMC-based estimation of this model, we here propose using the integrated nested Laplace approximation \citep[INLA,][]{rue2017bayesian} implemented in the \tt{INLA} package of the statistical software \tt{R}. It provides fast and ``off-the-shelf'' Bayesian inference for logistic regression models such as \eqref{eq:logreg}. We obtain the following parameter estimates, with $95\%$ credible intervals in parentheses: 
$$\hat{\beta}_{E,0}=-2.4\ (-2.8,-2.0), \quad \hat{\rho}_E=0.76\ (0.63,0.85),\quad \hat{\sigma}_E^2=2.1\ (1.1,3.5).$$ 
The estimated negative value of $\beta_{E,0}$ indicates that there is a higher probability for non-extreme events, as expected, while the estimated value of $\rho_E$ suggests that there is some non-negligible temporal persistence of zeros and ones. Figure \ref{fig:pfitted} shows the fitted probability values for a selection of $4$ years ($1,2,9,10$). The fitted models suggest relatively strong temporal autocorrelation: an extreme event at time $t$ entails an extreme event at time $t+1$ with probability substantially above average.

\subsection{Modeling extreme precipitation intensities}
 \label{sec:gammaGammaAppl}
  We now return to the modeling of (non-zero) precipitation intensities, and we fit the gamma-gamma model \eqref{eq:hierarchicalgammagamma} with latent Gaussian copula to the time series of selected extreme events. The model structure and data dimension are similar to those considered in the simulation study in \S\ref{sec:simulation}. We use standardized latitude, longitude, and altitude as covariates in the model. The dataset of extreme events is composed of $d=150$ sites and  $n=181$ days, leading to  $nd=27,150$ (spatially correlated) latent variables in the model. The minimum and maximum distances between the selected sites are $6$km and $819$km, receptively. We use $20$ sites for prediction and model validation (where the data are treated as completely missing). The remaining $130$ sites are used for model fitting. The proportion of missing observations at each of the $130$ training sites vary from $1\%$  to $10\%$, for an average of $3\%$. Our goal is also to compare how our model performs at different marginal thresholds, in order to assess the flexibility of our ``sub-asymptotic'' modeling approach, and we consider empirical quantiles at three moderately large probability levels, namely $85\%$, $90\%$ and $95\%$. Notice that while zero precipitation values may still occur at some spatial sites during extreme events, which creates a point mass at the lower endpoint of the precipitation distribution, we here avoid the tricky explicit treatment of zeros by censoring low precipitation intensities. We fit four different spatial models to the precipitation events obtained in \S\ref{sec:extmocc}, namely:
 \begin{description}
  \item[D$1$ \label{itm:GammagammaCovScale}] Gamma-gamma model \eqref{eq:hierarchicalgammagamma} with standardized spatial covariates given by latitude, longitude and altitude included in the scale parameter $\alpha$.
  \item[D$2$ \label{itm:GammagammaCovScaleShape}] Gamma-gamma model \eqref{eq:hierarchicalgammagamma} with standardized spatial covariates given by latitude, longitude, and altitude included in both scale $\alpha$ and shape $\beta_2$ parameters.
  \item[D$3$ \label{itm:ExpgammaCovScale}] Exponential-gamma (i.e., GP unconditionally) model, akin to \citet{Bopp.Shaby:2017}, with standardized spatial covariates given by  latitude, longitude, and altitude included in the scale parameter $\alpha$. This model may be obtained from Model D$1$ by fixing $\beta_1=1$.
  \item[D$4$ \label{itm:ExpgammaCovScaleShape}] Exponential-gamma (i.e., GP unconditionally) model, akin to \citet{Bopp.Shaby:2017}, with standardized spatial covariates given by  latitude, longitude, and altitude included in both scale $\alpha$ and shape $\beta_2$ parameters. This model may be obtained from Model D$2$ by fixing $\beta_1=1$.
  \end{description}
   
\begin{figure}[t!]
  \includegraphics[scale=0.8]{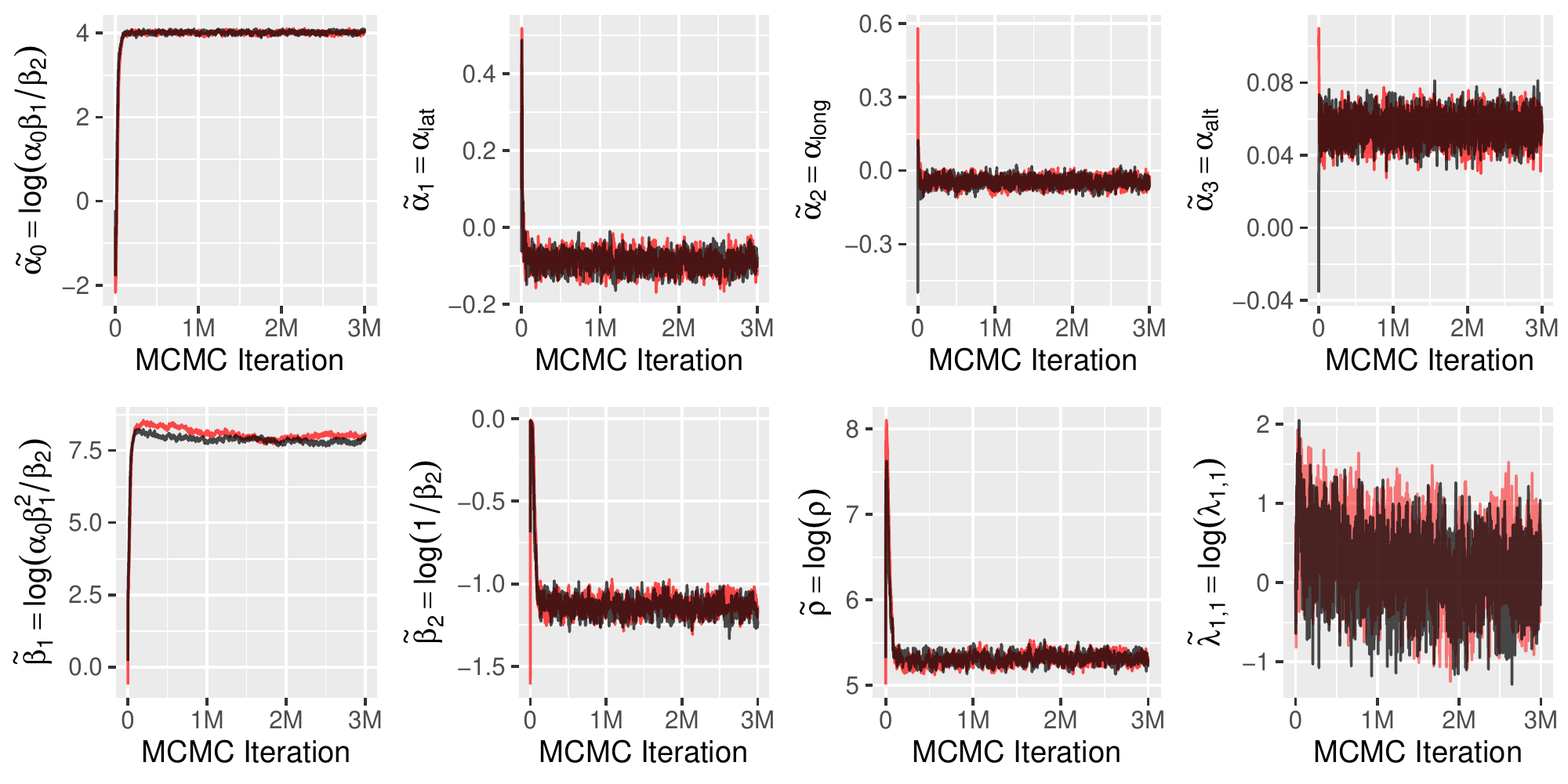}
   \caption{Trace plots of the $7$ hyperparameters, and of a selected latent variable (lower right plot) for model D$1$.  The marginal censoring threshold is here set to $90\%$. The red and black curves show two MCMC samples with different initial values. The total number of MCMC samples is $3,000,000$.}
    \label{fig:TracePlot}
\end{figure}

We fit all four models using the MCMC algorithm detailed in \S\ref{sec:bayesian}. We chose random initial values for all the hyperparameters, while we used the copula structure of latent variables defined in \eqref{eq:hierarchicalgammagamma} to generate initial values for the latent parameters. The trace plots displayed in Figure \ref{fig:TracePlot} show two MCMC chains with different initial values for all the hyperparameters and the latent parameter $\tilde{\lambda}_{1,1}=\log(\lambda_{1,1})$ (site $1$, day $1$) for model D$1$ with censoring threshold $90\%$. The behavior of the chains for the other latent variables is similar to $\tilde{\lambda}_{1,1}$. The Markov chains for the other models (D$2$, D$3$, D$4$) and for different censoring thresholds ($85\%$, $90\%$ $95\%$) are similar to those displayed in Figure \ref{fig:TracePlot}; see Figures~$9,10,13,15$, and $17$ in the Supplementary Material. The chains mix satisfactorily and converge to their stationary distribution after about $150,000$--$500,000$ iterations for most parameters. Figure~7 in the Supplementary Material shows trace plots of the tuning parameters $\tau_{\bm \lambda}$ and $\tau_{\bm \Theta}$ for the MALA and random walk proposals, respectively. The tuning parameters stabilize well before the end of the first burn-in phase of $750,000$ iterations, which illustrates the good performance of our adaptive MCMC algorithm. 

\begin{figure}[t!]
 \centering
    \includegraphics[width=\linewidth]{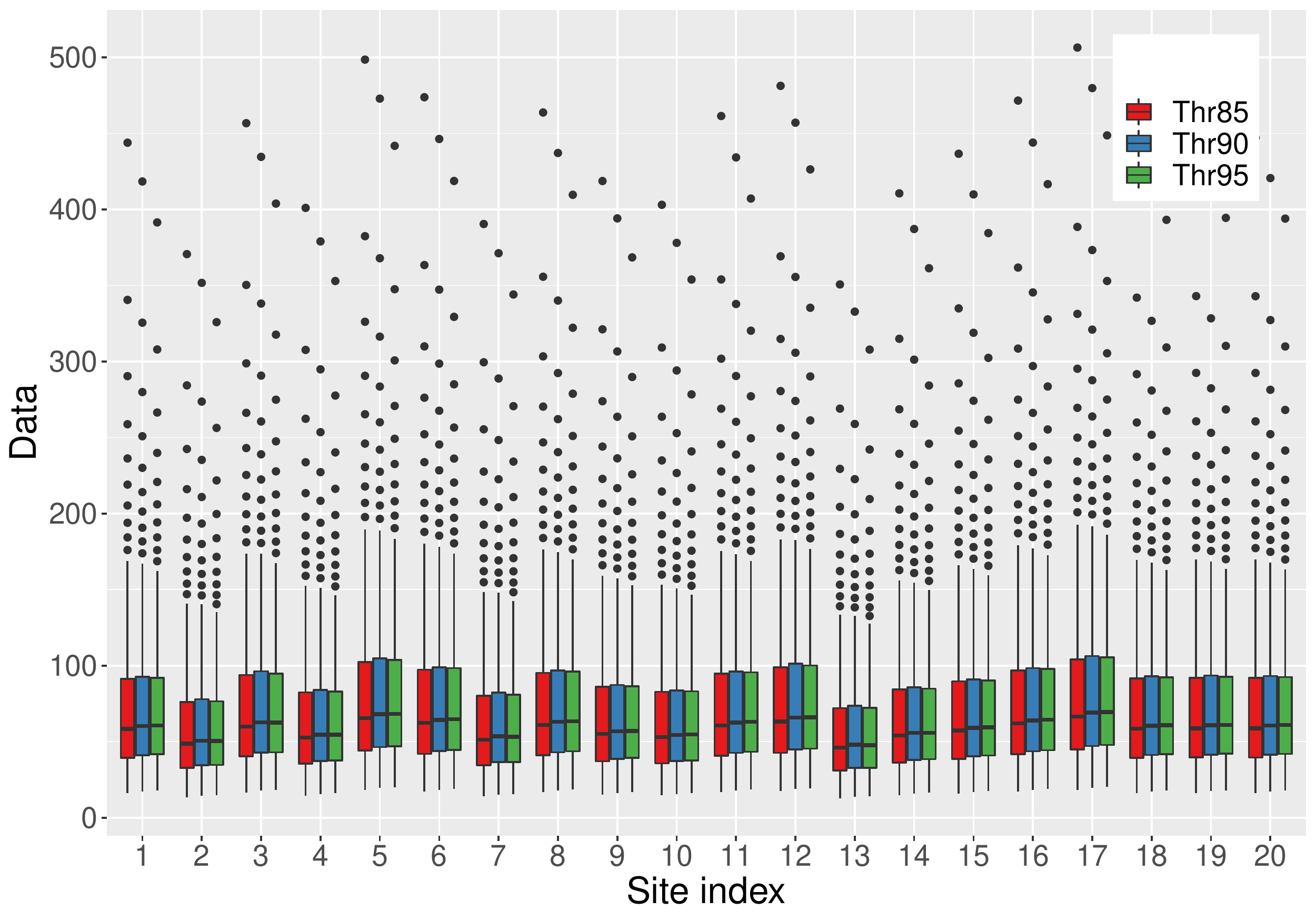}
    \caption{Boxplots of posterior predictive samples at all prediction sites from model D$1$ fitted to exceedances of the three different censoring thresholds (Thr) $85\%$, $90\%$, and $95\%$. The total number of MCMC samples is $3,000,000$, and the boxplots are based on the last $1,500,000$ MCMC samples, obtained after removing the first $1,500,000$ burn-in samples.}
    \label{fig:BoxPlotDiferentThreholds}
\end{figure}

In Table \ref{tab:twCRPS}, we compare models based on the continuous ranked probability score (CRPS) \citep{gneiting2007strictly} and the tail weighted continuous ranked probability score (twCRPS) \citep{lerch2017forecaster} for all the models (D$1$, D$2$, D$3$, and D$4$) and for different censoring thresholds ($85\%$, $90\%$, and $95\%$). While the CRPS is a proper scoring rule widely used to assess calibration and sharpness of probabilistic forecasts, the twCRPS is similar but focuses on the largest fraction of the data only (i.e., the upper tail). Based on the CRPS, model D$1$ appears to be the best amongst the four models considered, with much better scores than the exponential-gamma models (D$3$ and D$4$) and slightly better scores than the gamma-gamma model with covariates included both in the scale $\alpha$ and the shape $\beta_2$. However, based on the twCRPS, model D$2$ appears to be comparable to, yet still slightly better than model D$1$. Since model D$1$ is more parsimonious and easier to fit, has only a slight difference in twCRPS compared to model D$2$, and avoids issues of poor tail index predictions at unobserved sites (e.g., for unusual covariate values), we consider model D$1$ as our ``best'' model overall for our modeling extreme precipitation data over Germany. For brevity, we here only present the results for model D$1$; see the Supplementary Material for more details on the other models.
\begin{table}[t!]
		\small\addtolength{\tabcolsep}{10pt}
		\caption{CRPS and twCRPS values for all the candidate models, where the weight function in twCRPS is the Gaussian distribution function with variance $25$ and mean as the corresponding marginal threshold values. Lower values of CRPS and twCRPS are better. For each threshold and each diagnostic (CRPS/twCRPS), the best performance is highlighted in bold.}
		\centering

\begin{tabular}{c|ccc|ccc}

 & \multicolumn{3}{c|}{CRPS} & \multicolumn{3}{|c}{twCRPS}\\

\hline

 Model $\backslash$ Threshold& $85\%$ & $90\%$ & $95\%$ & $85\%$ & $90\%$ & $95\%$\\

\hline

D$1$ & \bf{35.32} & \bf{35.46} & \bf{35.56} & 9.17 & 6.77 & 3.95\\

D$2$ & 35.39 & 35.60 & 35.71 & \bf{9.05} & \bf{6.72} & \bf{3.90}\\

D$3$ & 39.55 & 40.21 & 42.07 & 9.23 & 6.79 & 3.95\\

D$4$ & 39.65 & 40.31 & 42.25 & 9.23 & 6.80 & 3.95\\

\end{tabular}
\label{tab:twCRPS}
\end{table}
Table \ref{tab:PostEstimateData} reports posterior mean estimates and two-sided $95\%$ credible intervals for all hyperparameters and some latent parameters for model D$1$. Interestingly, the results are fairly consistent across marginal censoring thresholds, indicating that our sub-asymptotic model can flexibly accommodate departure from the asymptotic GP distribution at finite levels; see also Figure \ref{fig:BoxPlotDiferentThreholds}, which compares boxplots of posterior predictive samples at all prediction sites for the three different censoring thresholds, $85\%$, $90\%$, and $95\%$.
\begin{table}[t!]
		\small\addtolength{\tabcolsep}{-3.5pt}
		\caption{Posterior mean (Post.\ mean) and $95\%$ Credible Interval (CI) of hyperparameters and of several latent parameters, reported for the censoring thresholds $85\%$, $90\%$, and $95\%$. The tail index parameter $\xi$ is equal to $1/\beta_2$. The total number of MCMC iterations is $3,000,000$. Estimations are based on the last $1,500,000$ MCMC samples, obtained after removing the first $1,500,000$ burn-in samples.}
		\centering
\begin{tabular}{c|cc|cc|cc}

 Threshold & \multicolumn{2}{c|}{$85\%$} & \multicolumn{2}{c|}{$90\%$} & \multicolumn{2}{c}{$95\%$}\\

\hline

 Parameter & Post.\ mean & $95\%$ CI & Post.\ Mean & $95\%$ CI & Post.\ Mean & $95\%$ CI\\

\hline

$\alpha_0$ & $1.16$ & $(1.04,1.30)$ & $1.27$ & $(1.16,1.38)$ & $1.39$ & $(1.23,1.51)$\\

$\alpha_{\text{lat}}$ & $-0.09$ & $(-0.12,-0.06)$ & $-0.09$ & $(-0.12,-0.06)$ & $-0.09$ & $(-0.13,-0.06)$\\

$\alpha_{\text{long}}$  & $-0.05$ & $(-0.07,-0.03)$ & $-0.05$ & $(-0.07,-0.02)$ & $-0.05$ & $(-0.08,-0.02)$\\

$\alpha_{\text{alt}}$  & $0.06$ & $(0.05,0.06)$ & $0.05$ & $(0.05,0.06)$ & $0.05$ & $(0.04,0.06)$\\

$\beta_1$ & $48.53$ & $(42.33,53.00)$ & $48.55$ & $(44.43,51.55)$ & $45.94$ & $(42.36,50.43)$\\

$\xi$ & $0.34$ & $(0.32,0.37)$ & $0.32$ & $(0.30,0.34)$ & $0.30$ & $(0.28,0.33)$\\

$\rho$ & $203.02$ & $(187.47,220.14)$ & $205.15$ & $(189.33,223.24)$ & $212.87$ & $(189.82,234.22)$\\

$\lambda_{1,1}$ & $0.85$ & $(0.51,1.31)$ & $1.26$ & $(0.67,2.08)$ & $1.02$ & $(0.54,1.72)$\\

$\lambda_{6,95}$ & $0.39$ & $(0.26,0.54)$ & $0.37$ & $(0.25,0.52)$ & $0.32$ & $(0.21,0.46)$\\

$\lambda_{56,95}$  & $0.71$ & $(0.45,1.06)$ & $0.62$ & $(0.41,0.92)$ & $0.80$ & $(0.43,1.36)$\\

\end{tabular}
\label{tab:PostEstimateData}
	\end{table}
As the conclusions are quite robust to the choice of the threshold, we now discuss the results for the $90\%$ threshold. The effect of the three covariates (latitude, longitude, altitude) is always significant as the $95\%$ credible intervals do not include $0$. This result demonstrates the importance of including suitable geographical information in the scale parameter $\alpha$ of the distribution. The estimates for latitude ($\hat{\alpha}_{\text{lat}}=-0.09$) and longitude ($\hat{\alpha}_{\text{long}}=-0.05$) indicate that the south-western part of Germany receives higher precipitation amounts than the north-eastern part---a pattern that is clearly perceptible in the mean precipitation plot in left panel of Figure~\ref{fig:SpatialMeanPlot}. Moreover, a clear positive effect of higher altitude on precipitation amounts arises with an estimate of ($\hat{\alpha}_{\text{alt}}=0.05$), which is also clear from the right panel of Figure~\ref{fig:SpatialMeanPlot}. The estimate of the shape parameter $\beta_1$ is around $49$ and shows a huge difference with respect to the generalized Pareto model with $\beta_1=1$. This finding substantiates our claim that extensions to the generalized Pareto distribution are useful for capturing complex data behavior at sub-asymptotic levels, and it confirms the comparison of models in Table~\ref{tab:twCRPS}. The estimated tail index $\hat{\xi}$ at the  $90\%$ censoring threshold is about $0.32$, which corresponds to quite heavy tails. The estimated  range parameter $\hat\rho$ of the exponential correlation function is  around $205$km, implying  a correlation of approximately $0.95$ \emph{at the latent} level between two sites separated by $10$km. This correlation  decreases to approximately $0$ between the two furthest sites. 

 \begin{figure}[t!]
 \centering
    \includegraphics[width=\linewidth]{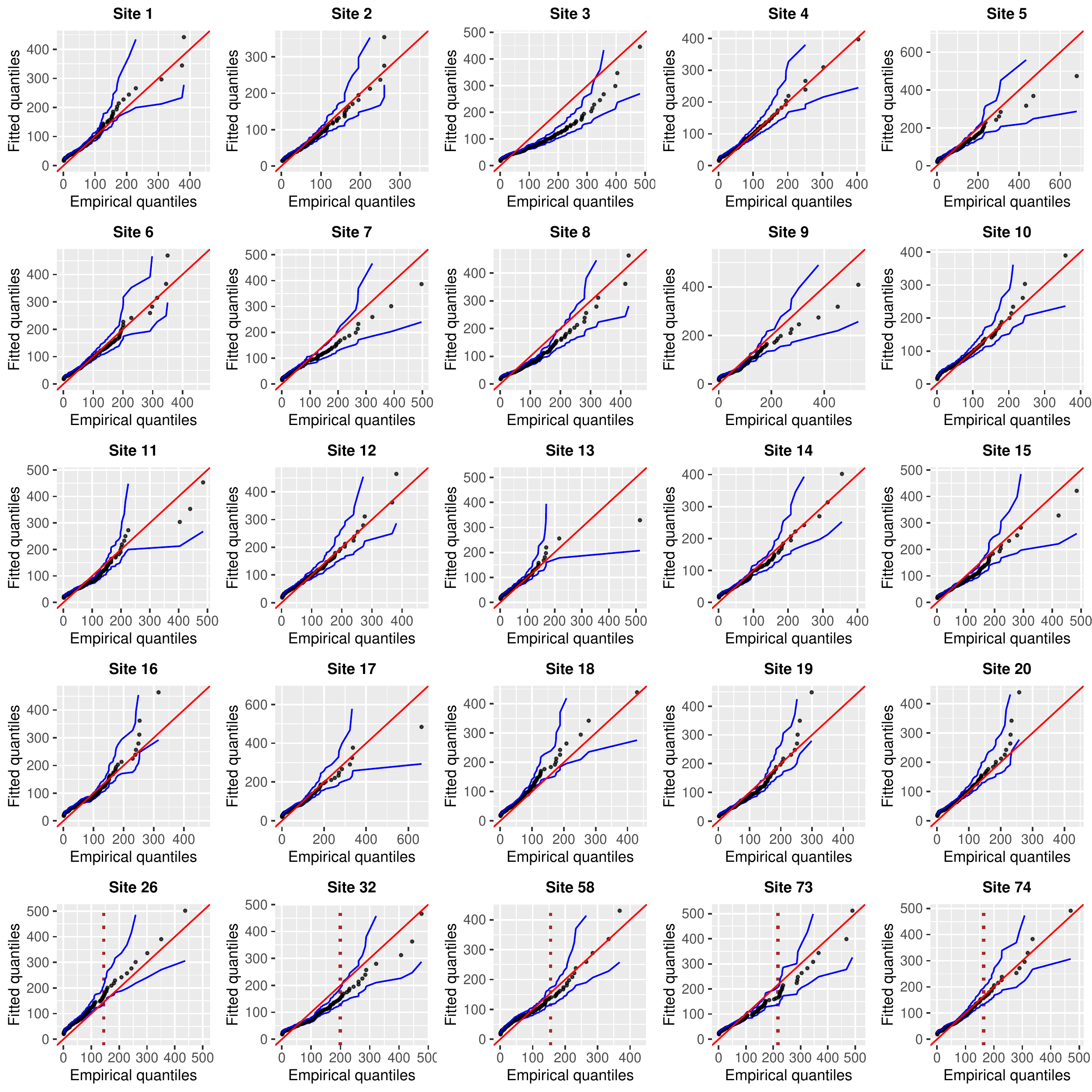}
    \caption{QQ-plots for all prediction sites for which the censoring threshold is set to $+\infty$ (top four rows), and some randomly chosen sites used for fitting the model (last row), based on model D$1$. Here, the marginal censoring thresholds are set to $90\%$. The vertical dotted brown lines in the last row indicates the threshold values. The QQ-plots are obtained by comparing the empirical data quantiles to the fitted quantiles calculated by plugging the posterior mean estimates into the unconditional $F$-distribution.}
    \label{fig:QQplotPredict}
\end{figure}


We now illustrate the spatial predictive performance of model D$1$ by QQ-plots and boxplots; see the Supplementary Material for diagnostics on the performance of the other models. To be concise, we here only discuss the results for the marginal censoring threshold $90\%$; see Figure \ref{fig:QQplotPredict}. The results for the other marginal thresholds are similar and reported in the Supplementary Material. The first four rows in Figure \ref{fig:QQplotPredict} correspond to the $20$ prediction sites, which are treated as fully missing, and the last row displays QQ-plots for $5$ randomly chosen sites used for fitting the model. We use the prediction sites to assess whether our fitted model appropriately predicts marginal distributions at unobserved sites. The model quantiles displayed in these QQ-plots are based on the unconditional $(\hat{\alpha} \hat{\beta}_1/\hat{\beta}_2)F_{2\hat{\beta}_1,2\hat{\beta}_2}$ distribution, where $\hat{\alpha}$, $\hat{\beta}_1$ and $\hat{\beta}_2$ are posterior mean estimates of $\alpha$, $\beta_1$ and $\beta_2$ for each site, respectively. We obtain $95\%$ uncertainty bands in the QQ-plots using a parametric bootstrap procedure accounting for missing values. Overall, the spatial predictive performance of model D$1$ is quite good, except perhaps at the prediction site 3, where our model tends to underestimate the empirical quantiles. Similar results were obtained for model D$2$ and other censoring levels with a generally satisfactory performance from low to high quantiles, but results are much worse for the exponential-gamma models D$3$ and D$4$; see the Supplementary Material for details. Boxplots of posterior predictive distributions based on model D$1$ for all prediction sites (see Figure $8$ of the Supplementary Material), roughly correspond to the empirical distribution of the data available at these sites, though with a slightly larger variability as expected. This suggests that our model is able to adequately capture the natural variability of precipitation intensities at unobserved locations. Based on our graphical diagnostics, model D$1$ seems to be the best overall to predict the precipitation distribution at unobserved locations. Furthermore, our results illustrate how the spatial dependence structure assumed at the latent level in our model can help predict precipitation over space. 
 
In summary, we conclude that our model appropriately captures spatial trends and dependence patterns of precipitation intensities during the extended autumn season in Germany. Specifically, it provides useful probabilistic predictions at unobserved locations during extreme events. 
 
%
%
%
%
%
%
%
%
%
%
%
%
%
%

\section{Conclusion}
\label{sec:conclusion}
We have proposed several univariate models extending the generalized Pareto limit model for threshold exceedances, and studied their tail properties. The high flexibility of these models suggests that they are good candidates for threshold-based modeling of moderate to large  values. 
Our models are based on hierarchical constructions using latent processes for spatial dependence. In this way, they avoid the artificial and overly strong separation of marginal and dependence modeling often encountered in spatial extreme value analysis.

Our Bayesian estimation  approach bears witness of the power of the MALA for simulation-based estimation in cases where the dimension of the latent model is comparable to the number of observations. Its use with censored data for exceedance-based spatial extreme-value analysis  allows bypassing high-dimensional numerical integration efficiently. Hence, our assumed model structure coupled with a data augmentation approach can be efficiently exploited to fit complex models in relatively high dimensions to censored threshold exceedances. 

In our spatio-temporal precipitation data application, we have opted for a two-step modeling approach of extreme quantiles. Essentially, we first identify spatial extreme events, defined here as days with large spatially aggregated values at all locations, and then separately model their occurrences and intensities. More precisely, in the first step, we model the binary time series of occurrences ($1$'s) or non-occurrences ($0$'s) of spatial extreme events through logistic regression, and in the second step, we model the precipitation intensities of extreme events using a spatial hierarchical gamma-gamma model. This two-step approach has the benefit of reducing the number of latent variables included in our Bayesian model fitted in the second step, speeding up computations. Temporal dependence may be incorporated through covariate or random effect modeling in the logistic regression. In future research, it would be interesting to further extend the latent process of our spatial hierarchical model to take the spatio-temporal dependence of successive extreme precipitation intensities into account.

\section*{Acknowledgments}
The data that support the findings of this study are openly available from the European Climate Assessment \& Dataset project at https://www.ecad.eu/. This publication is based upon work supported by the King Abdullah University of Science and Technology (KAUST) Office of Sponsored Research (OSR) under Award No. OSR-CRG2017-3434.

\baselineskip 16pt

\bibliographystyle{CUP}
\bibliography{ref}

\newpage 

\baselineskip 26pt

\baselineskip 10pt

\end{document}

%% file: macros.tex
\def\frac#1#2{{\textstyle{#1\over#2}}}

\DeclareSymbolFont{AMSb}{U}{msb}{m}{n}
\DeclareMathSymbol{\Natural}{\mathbin}{AMSb}{"4E}
\DeclareMathSymbol{\Integer}{\mathbin}{AMSb}{"5A}
\DeclareMathSymbol{\Real}{\mathbin}{AMSb}{"52}
\DeclareMathSymbol{\Rational}{\mathbin}{AMSb}{"51}
\DeclareMathSymbol{\Imaginary}{\mathbin}{AMSb}{"49}
\DeclareMathSymbol{\Complex}{\mathbin}{AMSb}{"43} 
\DeclareMathSymbol{\Disk}{\mathbin}{AMSb}{"44} 
\def\bi{\begin{itemize}}
\def\ei{\end{itemize}}
\def\bd{\begin{description}}
\def\ed{\end{description}}
\def\ben{\begin{enumerate}}
\def\een{\end{enumerate}}
\def\bv{\begin{verbatim}}
\def\ev{\end{verbatim}}

\def\calS{{\mathcal{S}}}

\def\hat#1{{\widehat{#1}}}

\newcommand{\indep}{\perp\!\!\!\perp}

\def\pr{{\rm Pr}}
\def\Pr{\pr}

\def\2to{{\ {\buildrel 2\over \longrightarrow}\ }}

\def\iid{{\ {\buildrel \rm{iid}\over \sim}\ }}

\def\I1ton{{$I_1,\ldots,I_n$}}
\def\X1ton{{$X_1,\ldots,X_n$}}
\def\Y1ton{{$Y_1,\ldots,Y_n$}}
\def\Z1ton{{$Z_1,\ldots,Z_n$}}
\def\R1ton{{$R_1,\ldots,R_n$}}
\def\e1ton{{$e_1,\ldots,e_n$}}
\def\t1ton{{$t_1,\ldots,t_n$}}
\def\x1ton{{$x_1,\ldots,x_n$}}
\def\y1ton{{$y_1,\ldots,y_n$}}
\def\z1ton{{$z_1,\ldots,z_n$}}

\def\calS{{\mathcal{S}}}

\def\tt#1{{\texttt{#1}}}

%% file: teachingmacros.tex
\def\tt#1{{\texttt{#1}}}

%% file: packages.tex
\usepackage[english]{babel}
\usepackage{amsfonts}
\usepackage{amssymb}
\usepackage{graphicx}
\usepackage{enumerate}